\documentclass{emulateapj}

\usepackage{graphicx,times}
\usepackage{epstopdf, epsfig}
\usepackage{multirow}
\usepackage{amsmath,amssymb}
\usepackage{rotating}

\usepackage{CJKutf8}

\submitted{{Accepted by \it ApJ}}

\begin{document}

\title{The Hawaii SCUBA-2 Lensing Cluster Survey: Number Counts and Submillimeter Flux Ratios}

\shortauthors{Hsu et al.}

\author{Li-Yen Hsu \begin{CJK*}{UTF8}{bsmi}(徐立研)\end{CJK*}\altaffilmark{1}, Lennox L. Cowie\altaffilmark{1}, Chian-Chou Chen \begin{CJK*}{UTF8}{bsmi}(陳建州)\end{CJK*}\altaffilmark{2,3}, Amy J. Barger\altaffilmark{1,4,5}, and \\ Wei-Hao Wang \begin{CJK*}{UTF8}{bsmi}(王為豪)\end{CJK*}\altaffilmark{6}}
\altaffiltext{1}{Institute of Astronomy, University of Hawaii, 2680 Woodlawn Drive, Honolulu, HI 96822, USA}
\altaffiltext{2}{Center for Extragalactic Astronomy, Department of Physics, Durham University, South Road, Durham DH1 3LE, UK}
\altaffiltext{3}{Institute for Computational Cosmology, Durham University, South Road, Durham DH1 3LE, UK}
\altaffiltext{4}{Department of Astronomy, University of Wisconsin-Madison, 475 North Charter Street, Madison, WI 53706, USA}
\altaffiltext{5}{Department of Physics and Astronomy, University of Hawaii, 2505 Correa Road, Honolulu, HI 96822, USA}
\altaffiltext{6}{Academia Sinica Institute of Astronomy and Astrophysics, P.O. Box 23-141, Taipei 10617, Taiwan}

\begin{abstract}

We present deep number counts at 450 and 850 $\mu$m using the SCUBA-2 camera on the James Clerk Maxwell Telescope. We combine data for six lensing cluster fields and three blank fields to measure the 
counts over a wide flux range at each wavelength. Thanks to the lensing magnification, our measurements extend to fluxes fainter than 1 mJy and 0.2 mJy at 450 $\mu$m and 850 $\mu$m, respectively. Our combined data highly constrain the faint end of the number counts. Integrating our counts shows that the majority of the extragalactic background light (EBL) at each wavelength is contributed by faint sources with $L_{\rm IR} < 10^{12} L_{\odot }$, corresponding to luminous infrared galaxies (LIRGs) or normal galaxies. By comparing our result with the 500 $\mu$m stacking of $K$-selected sources from the literature, we conclude that the $K$-selected LIRGs and normal galaxies still cannot fully account for the EBL that originates from sources with $L_{\rm IR} < 10^{12} L_{\odot }$. This suggests that many faint submillimeter galaxies may not be included in the UV star formation history. We also explore the submillimeter flux ratio between the two bands for our 450 $\mu$m and 
850 $\mu$m selected sources. At 850 $\mu$m, we find a clear relation between the flux ratio and the observed flux. This relation can be explained by a redshift evolution, where galaxies at higher redshifts have higher luminosities and star formation rates. 
In contrast, at 450 $\mu$m, we do not see a clear relation between the flux ratio and the observed flux.

\end{abstract}

\subjectheadings{cosmology: observations|  galaxies: formation  |  galaxies: starburst  |  gravitational lensing: strong | submillimeter: galaxies }

\section{Introduction}

Understanding the cosmic star formation history is crucial to constrain models for galaxy formation and evolution. To fully understand star-forming galaxies in the distant universe, observations at different wavelengths are required. The discovery of the far-infrared (FIR) Extragalactic Background Light (EBL) demonstrated that there is a comparable amount of light absorbed by dust and re-radiated into the FIR as there is detected directly in the optical and ultraviolet (UV) \citep{Puget1996Tentative-detec,Fixsen1998The-Spectrum-of,Dole2006The-cosmic-infr}. By $z > 1$, that FIR light is redshifted into the submillimeter. The first few deep-field maps made with the Submillimeter Common-User Bolometer Array (SCUBA;  \citealt{Holland1999SCUBA:-a-common}) on the 15m James Clerk Maxwell Telescope (JCMT) resolved the EBL into distinct sources at high redshifts (e.g., \citealt{Smail1997A-Deep-Sub-mill,Barger1998Submillimetre-w,Hughes1998High-redshift-s}), now known as submillimeter galaxies (SMGs; reviews by \citealt{Blain2002Submillimeter-g,Casey2014Dusty-Star-Form}). These SMGs are dusty, star-bursting galaxies, many of which cannot be easily picked out in the rest-frame UV or optical samples due to their high extinction (e.g., \citealt{Barger2012Precise-Identif,Simpson2014An-ALMA-Survey-}). The Spectral and Photometric Imaging Receiver (SPIRE) on the \emph{Herschel Space Observatory} \citep{Pilbratt2010Herschel-Space-} has also been very useful for mapping large areas of sky under the {\it Herschel} Multitiered Extragalactic Survey (HerMES; \citealt{Oliver2012The-Herschel-Mu}) and the {\it Herschel}-ATLAS Survey \citep{Eales2010The-Herschel-AT} to varying depths at 250, 350, and 500 $\mu$m, discovering rare and isolated bright FIR sources that have sometimes been found to be high-redshift lensed SMGs (e.g., \citealt{Negrello2010The-Detection-o,Conley2011Discovery-of-a-,Wardlow2013HerMES:-Candida,Bussmann2015HerMES:-ALMA-Im}). Many studies have shown that SMGs contribute a significant fraction of the cosmic star formation at high redshifts (e.g., \citealt{Barger2000Mapping-the-Evo,Barger2012Precise-Identif,Barger2014Is-There-a-Maxi,Chapman2005A-Redshift-Surv,Wang2006A-Near-Infrared,Serjeant2008The-SCUBA-Half-,2011MNRAS.415.1479W,Casey2013Characterisatio}). Thus, it is of great importance to have a complete understanding of this population of galaxies hidden by dust to map the cosmic star formation history fully.

\begin{table*}
\caption{Summary of SCUBA-2 Observations}
\begin{center}
\begin{tabular}{ccccccccc}
\hline \hline	

 Field  & RA & Dec & Scan Mode & Weather & Exposure  & Survey Area\footnote{The total observed area that we used for source detection at 450 $\mu$m and 850 $\mu$m. The 450 $\mu$m data of MACS\,J1149, MACS\,J1423, CDF-N, and CDF-S are shallow and are not used for constructing the number counts. Note that for a cluster field, the effective area on the source plane would be smaller than the quoted value here.} & $\bar{\sigma}$\footnote{Average 1$\sigma$ sensitivity within the survey area at 450 $\mu$m and 850 $\mu$m. These are the noise values measured from the reduced images, and the effect of confusion noise is not included.}    \\
           &       &        &      &          &   (hr)    &  (arcmin$^2$)  &  (mJy beam$^{-1}$)    \\

\hline

A1689                              & 13 11 29.0  & -01 20 17.0     & CV DAISY &1+2     &  20.4+1.9 & [120.2,125.3] & [4.44,0.69] \\ 
A2390                              &  21 53 36.8 &  \,\,17 41 44.2 & CV DAISY &1+2+3 &  11.4+21.5+9.0 & [126.1,131.6] & [7.14,0.66] \\
A370                                &  02 39 53.1 & -01 34 35.0     & CV DAISY &1+2+3  &  24.0+1.5+7.0  & [121.5,125.9] & [5.46,0.73] \\
MACS\,J0717.5+3745     &  07 17 34.0 &  \,\,37 44 49.0 & CV DAISY &1+2+3     &  24.2+3.5+1.5 & [127.0,127.3] & [4.62,0.73] \\            
MACS\,J1149.5+2223     & 11 49 36.3  &  \,\,22 23 58.1 & CV DAISY &1+2+3 &  6.0+2.0+2.4 & [---,121.9] & [---,1.23] \\
MACS\,J1423.8+2404     & 14 23 48.3  &  \,\,24 04 47.0 & CV DAISY &1+2+3 &  9.0+8.5+1.6 & [---,123.5] & [---,0.97] \\
%\hline
CDF-N                              &  12 36 49.0 &  \,\,62 13 53.0 & CV DAISY+PONG-900 &1+2+3 &  12.0+46.3+13.8 & [---,429.8] & [---,1.44] \\
CDF-S                              &  03 32 28.0 &  -27 48 30.0 & CV DAISY+PONG-900&1+2+3 &  3.7+53.1+5.5 & [---,314.3] & [---,1.75] \\
COSMOS                         &  10 00 24.0 &  \,\,02 24 00.0 & PONG-900 &1 &  38.0 & [379.5,377.3] & [5.65,0.99] \\

\hline\hline 
 
\end{tabular}  
\label{table1}
\end{center}
\end{table*}

Considerable observational effort has been expended to determine the FIR number counts because they provide fundamental constraints on empirical models (e.g., \citealt{Valiante2009A-Backward-Evol,Bethermin2011Modeling-the-ev}) and semi-analytical simulations \citep{2013MNRAS.434.2572H,2013MNRAS.428.2529H,Cowley2015,Lacey2015A-unified-multi} for galaxy evolution. Many measurements of the submillimeter number counts were made with SCUBA \citep{Smail1997A-Deep-Sub-mill,Hughes1998High-redshift-s,Barger1999Resolving-the-S,Eales1999,Eales2000,Cowie2002Faint-Submillim,Scott2002The-SCUBA-8-mJy,Smail2002The-nature-of-f,Borys2003The-Hubble-Deep,Serjeant2003Submillimetre-o,Webb2003The-Canada-UK-D,Wang2004An-850-Micron-S,Coppin2006The-SCUBA-Half-,Knudsen2008Probing-the-sub,Zemcov2010Contribution-of}. Similar results have been obtained with other single-dish telescopes and instruments, such as {\it Herschel} \citep{Oliver2010HerMES:-SPIRE-g,Berta2011}, the Large APEX Bolometer Camera (LABOCA; \citealt{Siringo2009The-Large-APEX-}) at 870 $\mu$m on the Atacama Pathfinder Experiment (APEX; \citealt{Gusten2006The-Atacama-Pat,Weis2009The-Large-Apex-}), and the AzTEC camera \citep{Wilson2008The-AzTEC-mm-wa} at 1.1 mm on both the JCMT (e.g., \citealt{Perera2008,Austermann2009AzTEC-Millimetr,Austermann2010AzTEC-half-squa}) and the Atacama Submillimeter Telescope Experiment (ASTE, \citealt{Ezawa2004The-Atacama-Sub}; e.g., \citealt{Scott2010Deep-1.1mm-wave,Scott2012The-source-coun,Aretxaga2011AzTEC-millimetr,Hatsukade2011AzTEC/ASTE-1.1-}).

The biggest challenge for measuring the FIR number counts is the poor spatial resolution of single-dish telescopes. For example, the beamsize of the JCMT at 850 $\mu$m is $\sim$ 15$''$; for {\it Herschel}, it is 18$''$, 26$''$, and 36$''$ at 250 $\mu$m, 350 $\mu$m, and 500 $\mu$m, respectively. Poor resolution imposes a fundamental limitation, the confusion limit \citep{Condon1974Confusion-and-F}, preventing us from resolving faint sources that contribute the majority of the EBL. Another issue caused by the poor resolution is source blending. Interferometric observations (e.g., \citealt{Wang2011SMA-Observation,Barger2012Precise-Identif,Smolcic2012Millimeter-imag,Hodge2013An-ALMA-Survey-,Bussmann2015HerMES:-ALMA-Im,Simpson2015The-SCUBA-2-Cos}) and semi-analytical models \citep{2013MNRAS.434.2572H,2013MNRAS.428.2529H,Cowley2015} have shown that close pairs within the large beam sizes are common.

Observations of massive galaxy clusters can push the detection limits toward fainter sources, thanks to gravitational lensing effects (e.g., \citealt{Smail1997A-Deep-Sub-mill,Smail2002The-nature-of-f,Cowie2002Faint-Submillim,Knudsen2008Probing-the-sub,Johansson2011A-LABOCA-survey,Chen2013Faint-Submillim,Chen2013Resolving-the-C}), though the positional uncertainties still cause large uncertainties in the lensing amplifications and the intrinsic fluxes \citep{Chen2011Submillimeter-S}. Serendipitous detection obtained within the deep, high-resolution ($\sim$ 1$''$) imaging taken by the Atacama Large Millimeter/submillimeter Array (ALMA) has allowed several measurements of number counts at 870 $\mu$m \citep{Karim2013An-ALMA-survey-,Simpson2015The-SCUBA-2-Cos}, 1.1 mm \citep{Carniani2015The-Cosmic-Infr}, 1.2 mm \citep{Ono2014Faint-Submillim,Fujimoto2016ALMA-Census-of-} and 1.3 mm \citep{Hatsukade2013Faint-End-of-1.,Carniani2015The-Cosmic-Infr}. However, the small-scale clustering between the random detections and the main targets may bias the counts (e.g., \citealt{Oteo2016ALMACAL-I:-Firs}). Unbiased measurements of submillimeter and millimeter number counts with ALMA still require imaging large areas of the 
sky \citep{Hatsukade2016SXDF-ALMA-2-arc}.

The SCUBA-2 camera \citep{2013MNRAS.430.2513H} on the JCMT has made it possible to search for SMGs efficiently. It covers 16 times the area of the previous SCUBA camera and has the fastest mapping speed at 450 $\mu$m and 850 $\mu$m with the best spatial resolution at 450 $\mu$m (FWHM $\sim$ 7$\farcs$5) among single-dish FIR telescopes. To exploit the capability of SCUBA-2 and to construct a large sample of faint SMGs that dominate the contribution of the EBL at 450 $\mu$m and 850 $\mu$m, we are undertaking a SCUBA-2 program, the Hawaii SCUBA-2 Lensing Cluster Survey (Hawaii-S2LCS), to map nine massive clusters, including the northern five clusters in the {\it HST} Frontier Fields program. While our program is still ongoing, we have already detected 99 and 478 sources at $\geq$ 4\,$\sigma$ at 450 $\mu$m and 850 $\mu$m, respectively. The preliminary results were published in \cite{Chen2013Faint-Submillim,Chen2013Resolving-the-C}. The individual properties of our detected sources will be presented in another paper (Hsu et al. 2016, in preparation).

In this paper, we present the 450 $\mu$m and 850 $\mu$m number counts constructed based on the SCUBA-2 observations of six cluster fields, A1689, A2390, A370, MACS\,J0717.5+3745, MACS\,J1149.5+2223, and MACS\,J1423.8+2404. To constrain the bright-end counts, we also include data from three blank fields, CDF-N, CDF-S, and COSMOS. We combine our measurements from all these fields in order to explore the widest possible flux range. The paper is structured as follows. The details of the observations and data reduction are described in Section~\ref{sec:data}. In Section~\ref{sec:count}, we explain our methodology for constructing the number counts and present our results. We discuss our results and their implications in Section~\ref{sec:dis}. Section~\ref{sec:sum} summarizes our results. Throughout this paper, we assume the concordance $\Lambda$CDM cosmology with 
$H_0=70~\rm km~s^{-1}~Mpc^{-1}$, $\Omega_M=0.27$, and $\Omega_\Lambda=0.73$ \citep{2011ApJS..192...16L}.

\section{Observations and Data Reduction}\label{sec:data}

We combined all of our SCUBA-2 data taken between October 2011 and January 2015, as well as the archival data of A1689 (PI: Holland) and COSMOS (PI: Casey; \citealt{Casey2013Characterisatio}). We used the 
CV DAISY scan pattern to observe our cluster fields, which detects sources out to a radius of $\sim$ 6$'$ and therefore covers the strong lensing regions of the clusters. We also used the PONG-900 scan 
pattern on the two CDF fields in order to cover larger areas to find rarer bright sources. Most of our observations were carried out under band 1 (the driest weather; $\tau_{\rm 225GHz} < 0.05$) or band 2 
($0.05 < \tau_{\rm 225GHz} < 0.08$) conditions, but there are also data taken under good band 3 conditions ($0.08 < \tau_{\rm 225GHz} < 0.1$).  The archival data of A1689 and COSMOS were taken under band 1 
conditions with the CV DAISY and PONG-900 modes, respectively. We summarize the details of these observations in Table~\ref{table1}.

Following \cite{Chen2013Faint-Submillim,Chen2013Resolving-the-C}, we reduced the data using the Dynamic Iterative Map Maker (DIMM) in the SMURF package from the STARLINK software \citep{2013MNRAS.430.2545C}. DIMM performs pre-processing and cleaning of the raw data (e.g., down-sampling, dark subtraction, concatenation, flat-fielding), as well as iterative estimations to remove different signals from astronomical signal and noise. We adopted the standard ``blank field'' configuration file, which is commonly used for extragalactic surveys to detect low signal-to-noise point sources. We ran DIMM on each bolometer subarray individually for a given scan and then used the MOSAIC\_JCMT\_IMAGES recipe from the Pipeline for Combing and Analyzing Reduced Data (PICARD) to coadd the reduced subarray maps into a single scan map. 

We then flux calibrated each scan with the primary calibrator observed closest in time \citep{2013MNRAS.430.2534D}. These calibrators are all compact bright sources such as Uranus, CRL618, CRL2688, and 
Arp220. We first reduced these calibrators with the ``bright compact'' configuration file and compared the derived flux conversion factors (FCFs) with the standard values provided in the SCUBA-2 data reduction 
manual (491 Jy pW$^{-1}$ at 450~$\mu$m and 537 Jy pW$^{-1}$ at 850~$\mu$m, derived with the ``bright compact'' configuration file). The resulting FCF values we obtained match these standard values to 
within 10\%, confirming the reliability of the calibrators we used. We then reduced the calibrators again using the same method used for the science maps with the ``blank field'' configuration 
file to derive a new set of FCFs. The derived values are on average $\sim$ 16\% and 20\% higher than the standard FCFs at 450~$\mu$m and 850~$\mu$m, respectively. We applied these FCFs to the the science scans.

After each scan was reduced and flux calibrated, we used MOSAIC\_JCMT\_IMAGES again to combine all the products into the final maps. Finally, to maximize the detectability of point sources, we applied a matched 
filter to our maps using the PICARD recipe SCUBA2\_MATCHED\_FILTER. Before running the matched filter, the recipe convolved the maps with a broad Gaussian and subtracted these maps from the original maps in order 
to remove low spatial frequency structures. We adopted the default FWHM values for the broad Gaussian (20$''$ at 450~$\mu$m and 30$''$ at 850~$\mu$m). The processed point-spread function (PSF) used for matched-filtering is a Gaussian 
with a convolved broader Gaussian subtracted off, which gives a Mexican-hat-like wavelet\footnote{A matched filter using a two-component model of the JCMT beam to create PSFs has been adopted since the 2014A release of STARLINK 
software (Date Released: July 24, 2014). However, in this work, we used the older ``Hikianalia'' release to reduce and combine our data. The choice of the software does not affect the matched-filter subtraction in any significant way.}.

\section{Number Counts}\label{sec:count}

\subsection{Pure Noise Maps}

In order to estimate the number of fake sources contaminating the number counts that we measured from our science maps, we need to generate source-free maps with only pure 
noise for each of our fields. These maps are sometimes referred to as jackknife maps in the literature. Following \cite{Chen2013Faint-Submillim,Chen2013Resolving-the-C}, we subtracted two maps that were each produced by coadding roughly half of the flux-calibrated data. In doing so, the real sources are subtracted off, and the residual maps are source-free maps. We then rescaled the value of each pixel by a factor of $\sqrt{t_1 \times t_2}/(t_{1}+t_{2})$, with $t_1$ and $t_2$ representing the integration time of each pixel from the two maps. Finally, we applied the matched-filter with the same procedure for the science maps.

\subsection{Source Extraction}

We have shown in previous work that sources detected above a 4$\,\sigma$ level have a low contamination rate \citep{Chen2013Faint-Submillim,Chen2013Resolving-the-C}. However, for computing 
number counts, we can use a lower detection threshold where there are still significantly more true sources than false detections. In \cite{Chen2013Resolving-the-C}, we extracted sources down to 
$\sim 2\,\sigma$. However, here we use 3$\,\sigma$ as our detection threshold\footnote{Note that the confusion noise is not included in the flux errors of detected sources.}. We experimented with different detection
 thresholds and binning in order to extend the faint end of the counts while keeping good signal-to-noise ratio (S/N) throughout all the flux bins at the same time. We found that using 3$\,\sigma$ leads to better S/N in the number 
 counts than using lower thresholds, especially at the faint end.

We first generated the PSFs by averaging all the primary calibrators, the ones we used for deriving the FCFs. Following the methodology of source extraction in \cite{Chen2013Faint-Submillim,Chen2013Resolving-the-C}, we identified the pixel with the maximum S/N, subtracted this pixel and its surroundings using the PSF centered and scaled at the position and value of this pixel, and then searched for the next maximum S/N. We iterated this process until the 3\,$\sigma$ threshold was hit. We ran the source extraction on both the science maps and and the pure noise maps. The ratio of the total number of sources from the pure noise maps and from the science maps ($= N_{\rm false}/N_{\rm total}$) is 
$\sim$ 55\% (22\%) at 450 (850) $\mu$m. The effect of false sources is subtracted in the computation of number counts, as we will describe in Section~\ref{sec:delensing} and \ref{sec:simulation}.

\begin{figure}
\begin{center}    
\includegraphics[width=7.5cm]{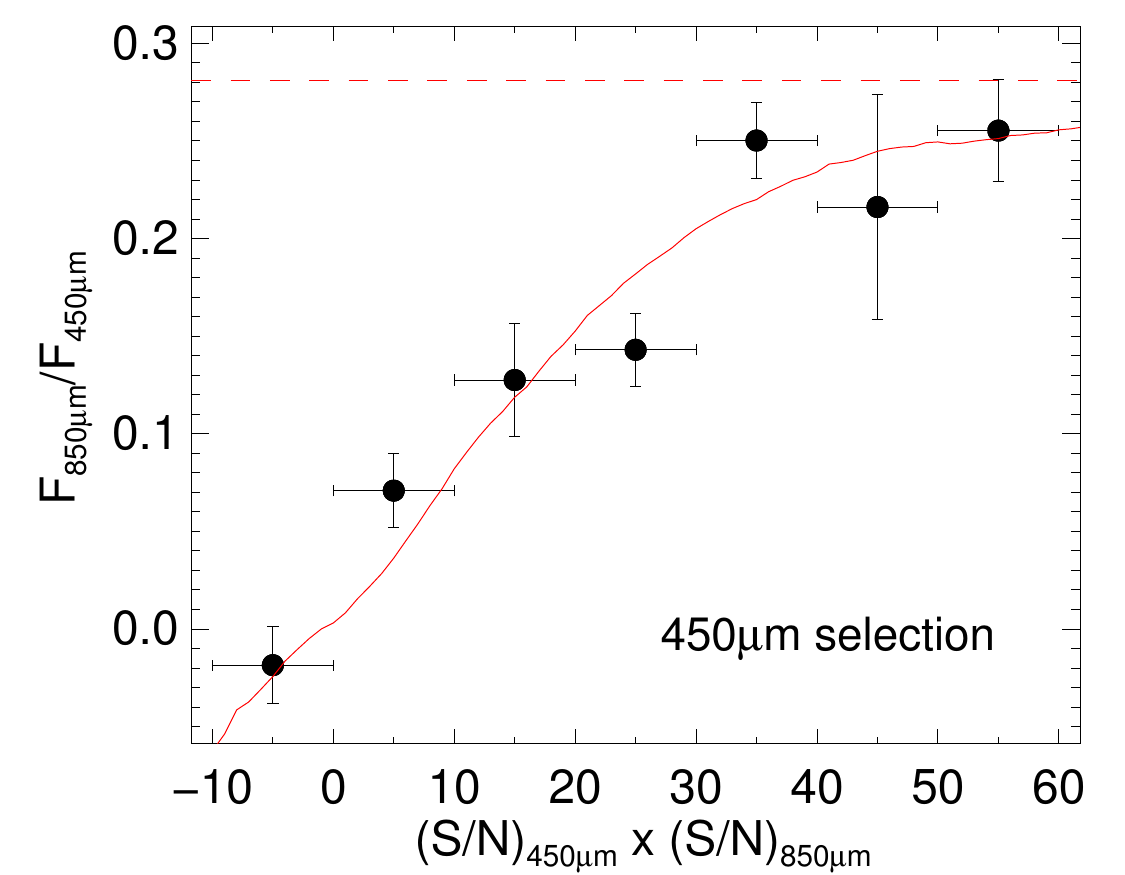}   
\includegraphics[width=7.5cm]{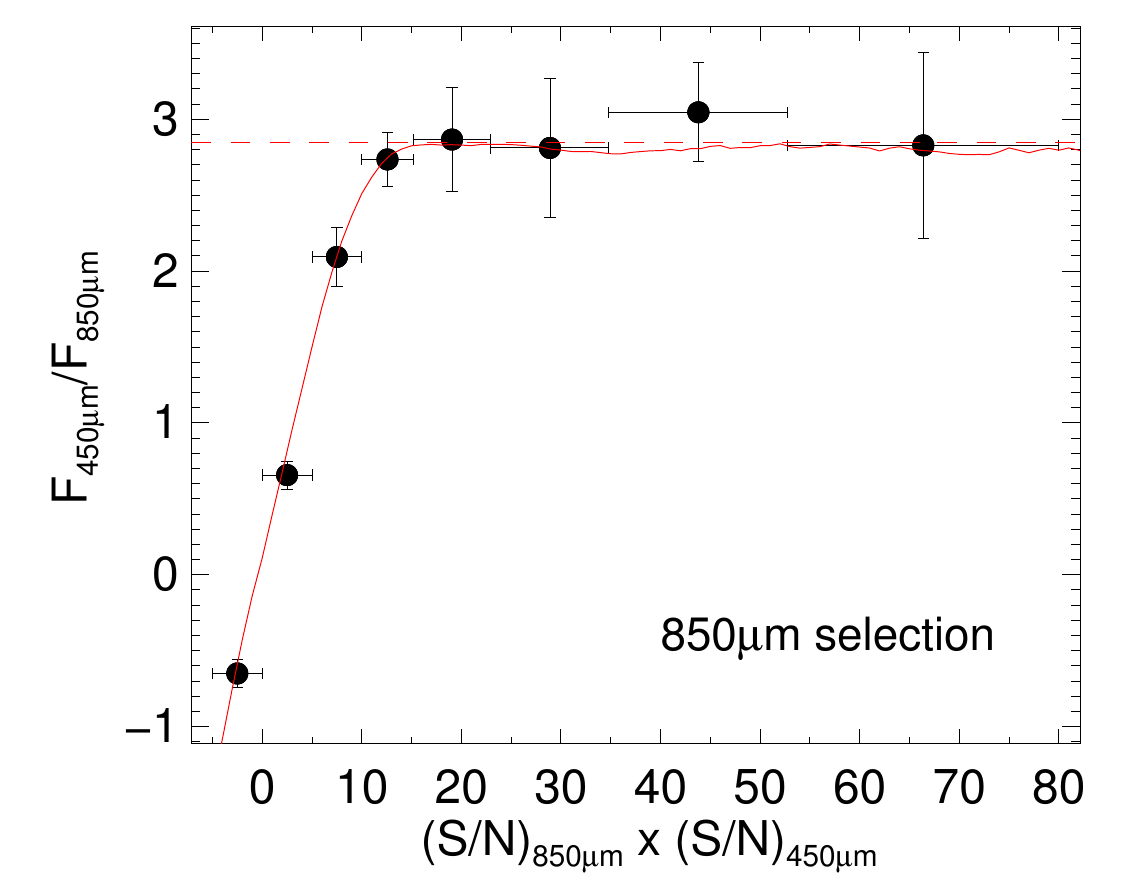}  
\caption{850$\mu$m-to-450$\mu$m (upper) and 450$\mu$m-to-850$\mu$m (lower) flux ratios against (S/N)$_{450\mu{\rm m}} \times$ (S/N)$_{850\mu{\rm m}}$ for 4$\,\sigma$ detected sources in the cluster fields 
at 450 $\mu$m and 850 $\mu$m, respectively. In each bin, we took the median of the flux ratios and calculated the error using bootstrapping. Note that for 850 $\mu$m detected sources we use logarithmic 
bins at (S/N)$_{450\mu{\rm m}} \times$ (S/N)$_{850\mu{\rm m}} > 10$ in order to improve the number statistics. We ran simulations in which we populated the source-free pure noise maps with sources with 
constant flux ratios. Red dashed lines are the input constant flux ratios (0.281 and 2.85) of the simulations, and the red solid lines are what we measured from the simulated maps, which show 
good agreement with our measurements from the science maps.}  
\end{center}
\label{figure1}
\end{figure}

\subsection{Submillimeter Flux Ratios and Redshift Estimates}\label{sec:color}

In order to obtain the intrinsic flux of a lensed source, both the lens model of the cluster and the redshift of the source are required. However, since we do not have redshift measurements for individual sources, we simply 
adopted estimated median redshifts of the lensed 450 $\mu$m and 850 $\mu$m sources to compute our de-lensed number counts. Note that what we need to estimate are the ``observed median redshifts'' of the 
lensed populations, which would be higher than the real median redshifts of the distributions (e.g., \citealt{Weis2013ALMA-Redshifts-}). This is because sources at higher redshifts have higher probability of being lensed and generally have higher lensing magnifications, causing a selection bias (e.g., \citealt{Hezaveh2011Effects-of-Stro}). We leave the discussion of the blank-field sources and their redshift distributions to Section~\ref{sec:redshift}.

We estimated the two median redshifts by exploring the flux ratios between 450 $\mu$m and 850 $\mu$m for all the 4$\,\sigma$ detected sources in the cluster fields. At 450 (850) $\mu$m, we took the flux and position of a detected source and then measured the flux value at the same position on the 850 (450) $\mu$m map. In Figure~1, we plot the 850$\mu$m-to-450$\mu$m and 450$\mu$m-to-850$\mu$m flux ratios against the product of S/N at 450 $\mu$m and at 850 $\mu$m. In each bin of (S/N)$_{450\mu{\rm m}} \times$ (S/N)$_{850\mu{\rm m}}$, we took the median of the flux ratios and calculated the error using bootstrapping. We can see that the flux ratio increases with increasing S/N product and then flattens. The lower (negative) measured flux ratios at lower (negative) (S/N)$_{450\mu{\rm m}} \times$ (S/N)$_{850\mu{\rm m}}$ are a result of the mismatch between the positions of the 450 $\mu$m and 850 $\mu$m flux peaks due to lower S/N. We compared the measured flux ratios with what we measured from simulated maps, which were produced by populating the pure noise maps with sources with constant flux ratios. A detailed description of how we performed such simulations is left to the Appendix. In Figure~1, red dashed lines are the input constant flux ratios of the simulations, and the red solid lines are what we measured from the simulated maps, which show good agreement with our measurements from the science maps. We therefore conclude that the values of the two dashed lines correspond to the median flux ratios of the 450 $\mu$m and 850 $\mu$m selected populations in the cluster fields.

To convert flux ratios to redshifts, we assumed a modified blackbody spectral energy distribution (SED) of the form $S_{\nu} \propto (1 - e^{-\tau(\nu)})B_{\nu}(T)$, where $\tau(\nu) = (\nu / \nu_0)^{\beta}$ and $\nu_0 =3000$ GHz. Assuming 
$\beta = 1.5$, we determined the redshifts from our estimated median flux ratios for a dust temperature of 30 K, 40 K, or 50 K. The corresponding redshifts for the 450 $\mu$m sources are $z \sim$ 1.5, 2.2, or 2.8. At 850 $\mu$m, we obtained 
$z \sim$ 2.0, 2.8, or 3.5. The final number counts shown in this paper are based on source plane redshifts of 2.2 and 2.8 for 450 $\mu$m and 850 $\mu$m, respectively. We chose these values because they are the central values of the different SED models used. However, we will show in Section~\ref{sec:result} that using $z = 1.5$, 2.8 at 450 $\mu$m and $z = 2.0$, 3.5 at 850 $\mu$m does not change our results significantly, and that the computation of the number 
counts is not sensitive to the adopted source plane redshifts.

\subsection{De-lensed Raw Number Counts}\label{sec:delensing}

To compute the de-lensed, differential number counts, we corrected all the source fluxes in the cluster fields using the publicly available software {\sc LENSTOOL} \citep{Kneib1996Hubble-Space-Te}, which allows us to generate magnification maps with the angular sizes of our SCUBA-2 maps. We therefore used the lens models from the {\sc LENSTOOL} developers (CATS team) for A1689 \citep{Limousin2007Combining-Stron}, A2390
\citep{Richard2010LoCuSS:-first-r}, MACS\,J1423.8+2404 \citep{Limousin2010MACS-J1423.8240}, and the three Frontier Fields (Hubble Frontier Field archive\footnote{https://archive.stsci.edu/pub/hlsp/frontier/}). For each source from a science or pure noise map, we calculated its number density by inverting the detectable area, which is the area in which this source can be detected above the 3\,$\sigma$ threshold. For a source in a cluster field, the detectable area is defined on the source plane. We then computed the number counts by summing up the number densities of the sources in each flux bin with errors based on Poisson statistics \citep{Gehrels1986Confidence-limi}. Finally, we subtracted the counts of the pure noise maps from the counts of the science maps to produce the pure source counts.

While the discrepancy in the magnifications between different lens models can be a factor of a few at the cluster center, the effect on the measured number counts is not significant. This is the same as the effect caused by the different source
plane redshifts, as we discussed in Section~\ref{sec:color}. Although there are uncertainties in the lens models, the source plane redshifts, and the positions of the submillimeter sources, the de-lensed flux and detectable area 
of a source are directly related, causing little change in the slope and normalization of the measured number counts.

\begin{figure}
\begin{center}    
\includegraphics[width=7.5cm]{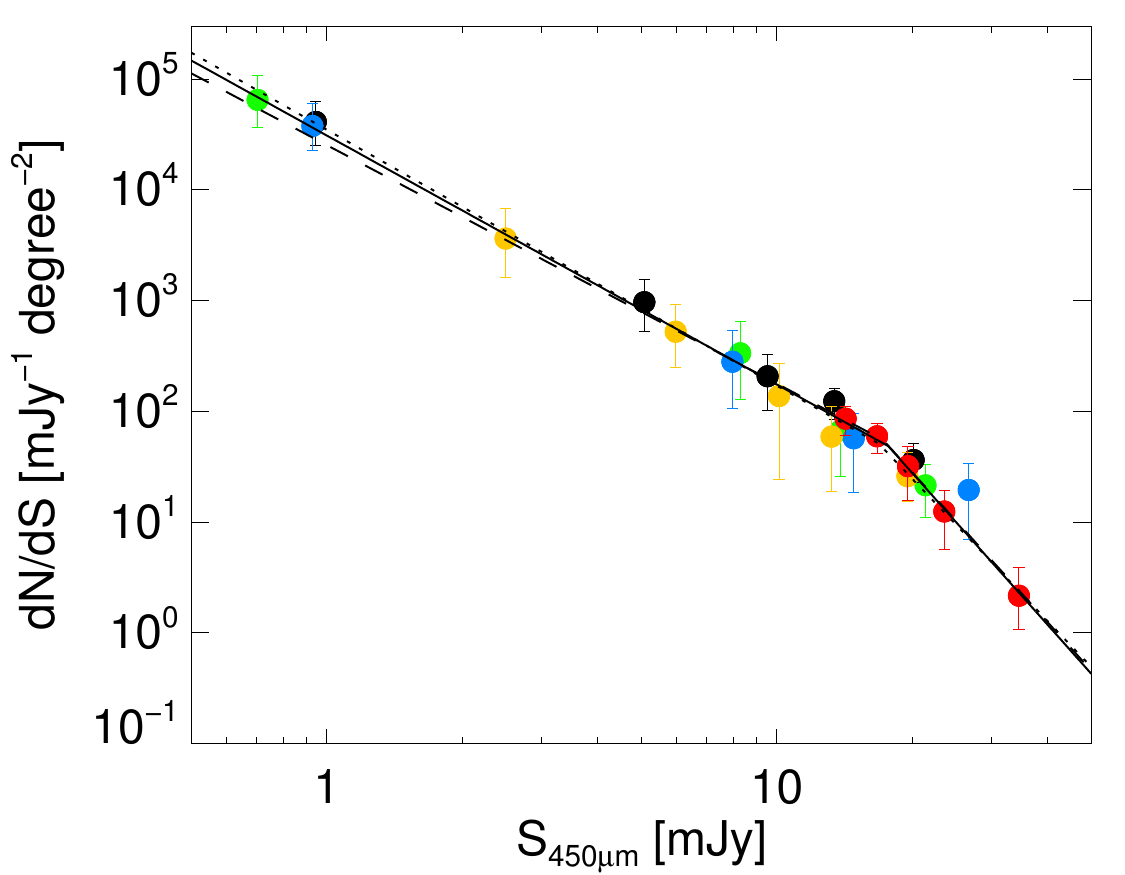}   
\includegraphics[width=7.5cm]{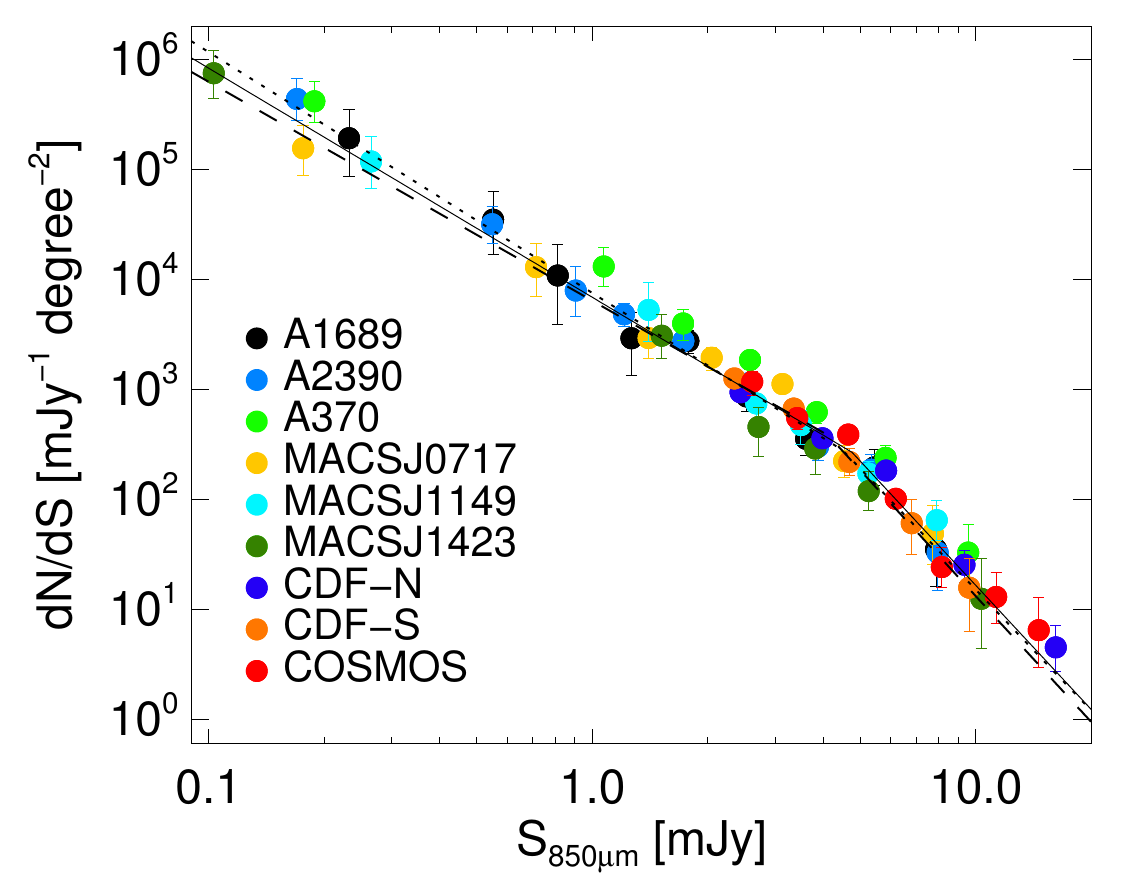}  
\caption{Differential number counts for all the fields at 450 $\mu$m (upper) and 850 $\mu$m (lower), assuming source plane redshifts of 2.2 (450 $\mu$m) and 2.8 (850 $\mu$m) for de-lensing. Solid lines are the best-fit 
broken power law models. Dotted lines are the best-fit models to the number counts computed with source plane redshifts of 1.5 and 2.0. Dashed lines are the best-fit models to the number counts computed with source 
plane redshifts of 2.8 and 3.5. We do not show the counts using these different source plane redshifts for clarity.}   
\end{center}
\label{fig:figure2}
\end{figure}

\begin{figure}
\begin{center}    
\includegraphics[width=8cm]{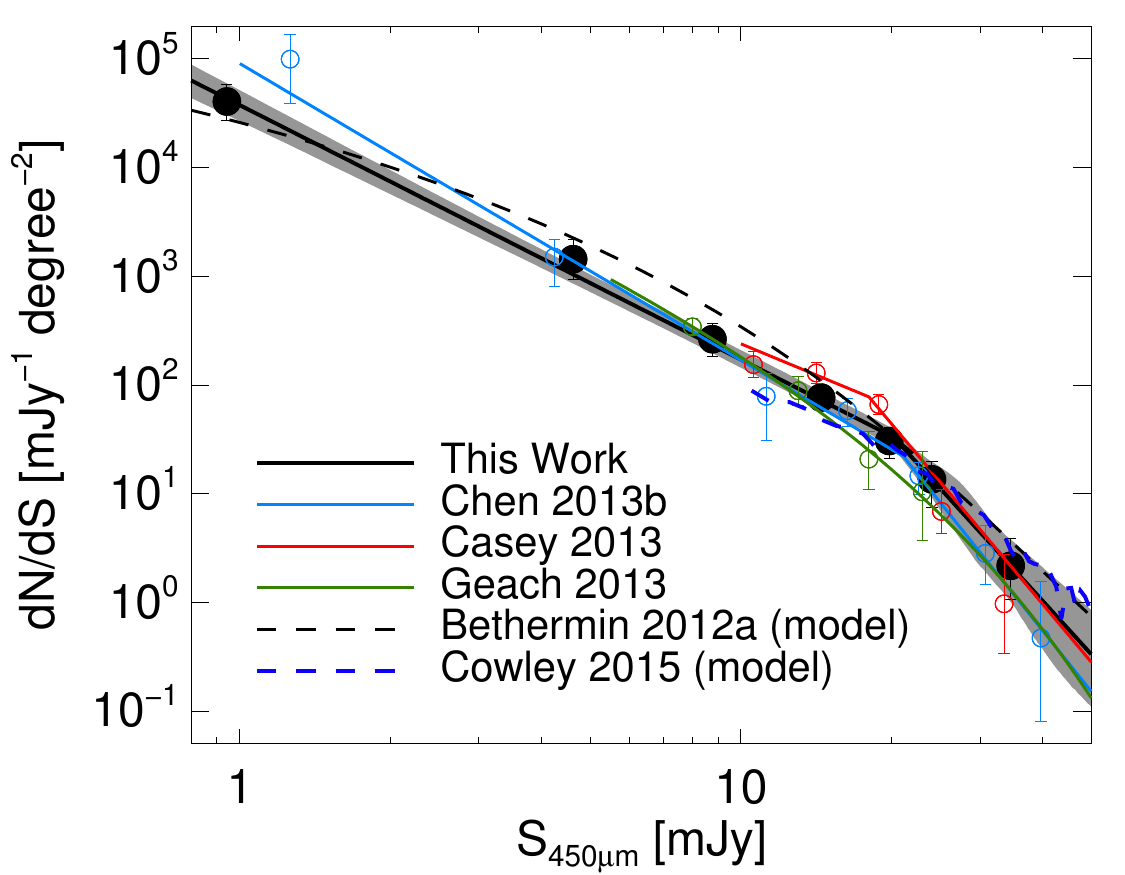} 
\includegraphics[width=8cm]{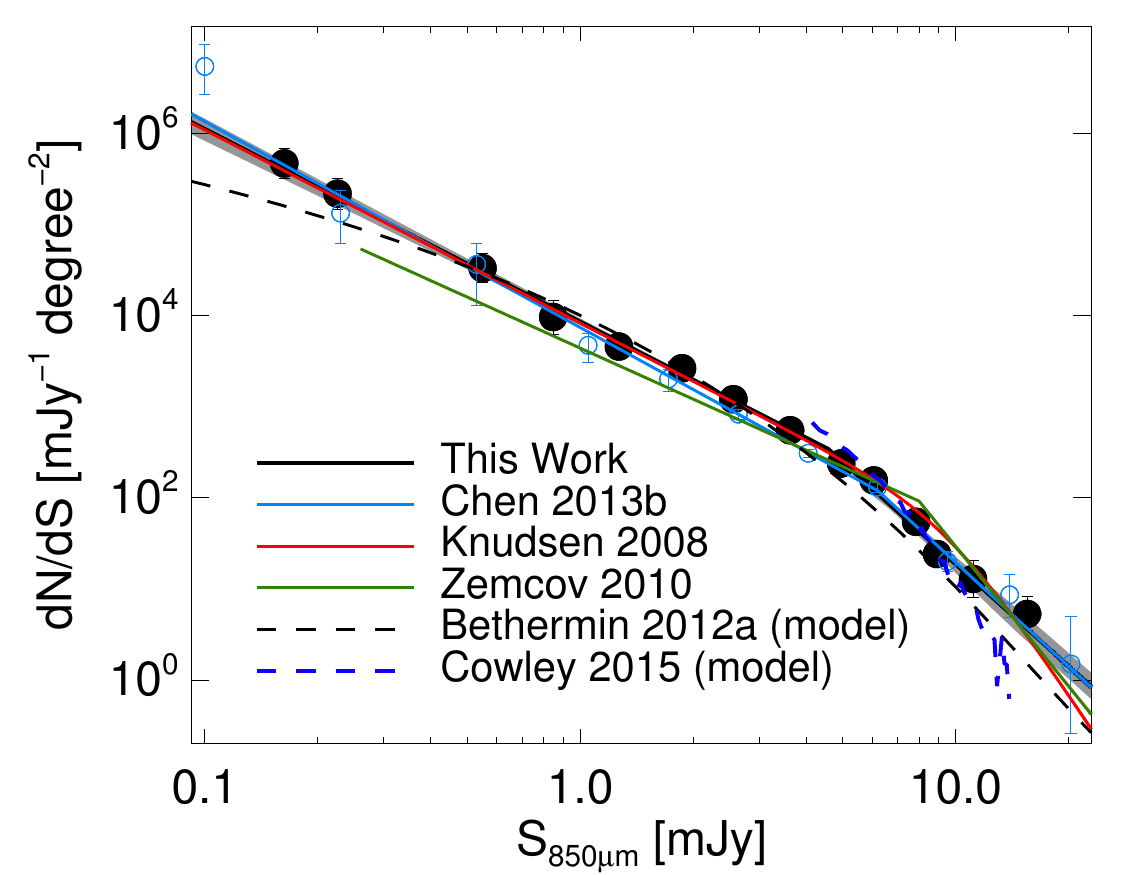}  
\caption{Combined differential number counts from all the available fields. Solid black lines are the best-fit broken power law models with 1\,$\sigma$ error regions in gray shading. The values of the counts and the best-fit parameters are summarized in Table \ref{table2} and \ref{table3}, respectively. In both panels, we show the best-fit count models for a few other observational results as colored solid lines, and the model predictions from \cite{Bethermin2012A-Unified-Empir} and \cite{Cowley2015} as black and purple dashed lines, respectively. For the observational results from the literature, the measured number counts are shown as colored circles if their values are available from these papers. Blue lines and circles are from \cite{Chen2013Resolving-the-C}. At 450 $\mu$m, the results from \cite{Casey2013Characterisatio} and \cite{2013MNRAS.432...53G} are shown in the upper panel as red and green lines/circles, respectively. At 850 $\mu$m, two count models from SCUBA cluster surveys are plotted in the lower panel as red \citep{Knudsen2008Probing-the-sub} and green \citep{Zemcov2010Contribution-of} lines. Note that the original models from \cite{Cowley2015} are cumulative, and we 
converted these models to differential counts using flux intervals of 1 mJy and 0.2 mJy at 450 $\mu$m and 850 $\mu$m, respectively. Using smaller flux intervals would make the resulting differential counts less smooth.}  
\end{center}
\label{figure3}
\end{figure}

\subsection{Simulations and Corrected Number Counts}\label{sec:simulation}

The pure source counts we computed above, however, still do not represent the true underlying submillmeter populations because the fluxes of the sources are boosted and there is incompleteness. The cause of the flux boost is the statistical 
fluctuations of the flux measurements for flux-limited observations, known as the Eddington bias \citep{Eddington1913On-a-formula-fo}. Following \cite{Chen2013Faint-Submillim,Chen2013Resolving-the-C}, we ran Monte Carlo simulations to estimate the underlying count model at each wavelength. We first generated a simulated map by randomly populating sources in the pure noise map, drawn from an assumed model and convolved with the PSF. The count model 
we used is in the form of a broken power law
\begin{equation}                                                               
\frac{dN}{dS}=\left\{\begin{matrix}
N_0\left ( \frac{S}{S_0} \right )^{-\alpha} ~~{\rm if}~S\leq S_0 \\ 
N_0\left ( \frac{S}{S_0} \right )^{-\beta} ~~{\rm if}~S > S_0
\end{matrix}\right.
\end{equation}  
For the cluster fields, we populated the sources in the source plane and projected them onto the image plane using {\sc LENSTOOL}.

For each simulated map we reran our source extraction and computed the recovered counts using the same method and flux bins used for the science map. We repeated the simulation 50 times for each input model and 
then averaged the recovered counts from these simulations. In order to measure the actual counts we adopted an iterative procedure. Using the ratios between the averaged recovered counts and the input counts, we renormalized 
the observed raw counts in each bin from the science map. We then did a $\chi^2$ fit to the corrected observed counts using a broken power law. This fit was then used as the input model for the next iteration. We continued until 
the corrected counts matched the corrected counts of the previous iteration within 1\,$\sigma$ throughout all the flux bins. It took only two or three iterations to converge for each field.

\subsection{Results}\label{sec:result}
We show the corrected number counts for all the fields together at both 450 $\mu \rm{m}$ and 850 $\mu \rm{m}$ in Figure~2. Thanks to the lensing magnification, we are able to detect counts down to fluxes fainter 
than 1 mJy and 0.2 mJy in several fields at 450 $\mu$m and 850 $\mu$m, respectively. The solid lines represent the best-fit broken power law models for the counts. In each panel, we also 
show the best-fit model for the counts computed with a lower source plane redshift ($z=1.5$ at 450 $\mu \rm{m}$ and $z=2.0$ at 850 $\mu \rm{m}$) with the dotted line and the best-fit model to the counts computed with 
a higher source plane redshift ($z=2.8$ at 450 $\mu \rm{m}$ and $z=3.5$ at 850 $\mu \rm{m}$) with the dashed line. We can see that the results are not very sensitive to the assumed redshifts. 

In order to better constrain our count model at each wavelength, we combined the counts from all the fields. The results are shown in Figure~3, which are weighted averages of the corrected counts from each field (black circles). 
We assigned a weight for each flux bin of a field in the following way. For each field, we used the final count model we obtained (Section~\ref{sec:simulation}) to run the same simulation 50 times and obtained the 1\,$\sigma$ scatter of the 
recovered counts in each flux bin. We then normalized this scatter by the average of the recovered counts. The inverse square of the scatter is adopted as the weight. We also show various results from the 
literature. The combined number counts and the best-fit parameters of the models are summarized in Table~\ref{table2} and Table~\ref{table3}.

\begin{table}
\caption{The Combined Differential Number Counts at 450 $\mu{\rm m}$ and 850 $\mu{\rm m}$}
\begin{center}
\begin{tabular}{cccc}
\hline \hline	

 $S_{450\mu \rm{m}}$  & $dN/dS$ & $S_{850\mu \rm{m}}$ & $dN/dS$    \\
  (mJy)      &   (mJy$^{-1}$~deg$^{-2}$)  &  (mJy)    &  (mJy$^{-1}$~deg$^{-2}$)  \\

\hline

  \multirow{2}{*}{0.94} &    \multirow{2}{*}{40579$\substack{+17834 \\ -13496}$} & \multirow{2}{*}{0.16} &    \multirow{2}{*}{468599$\substack{+215679 \\ -147442}$}\\
      &  &   & \\  
  \multirow{2}{*}{4.63} &   \multirow{2}{*}{1438$\substack{+741 \\ -510}$} & \multirow{2}{*}{0.23} &    \multirow{2}{*}{217910$\substack{+105487 \\ -70715}$}\\
        &  &   & \\  
  \multirow{2}{*}{8.77} &   \multirow{2}{*}{263.9$\substack{+105.7 \\ -79.2}$} & \multirow{2}{*}{0.55} &    \multirow{2}{*}{33138$\substack{+15129 \\ -9975}$}\\
        &  &   & \\  
  \multirow{2}{*}{14.45} &  \multirow{2}{*}{76.58$\substack{+18.52 \\ -15.96}$} & \multirow{2}{*}{0.85} &    \multirow{2}{*}{9650$\substack{+4984 \\ -3444}$}\\
        &  &   & \\  
  \multirow{2}{*}{19.76} &  \multirow{2}{*}{30.53$\substack{+10.02 \\ -9.39}$} & \multirow{2}{*}{1.27} &    \multirow{2}{*}{4576$\substack{+1114 \\ -826}$}\\
        &  &   & \\  
  \multirow{2}{*}{24.05} &  \multirow{2}{*}{13.55$\substack{+6.17 \\ -6.09}$} & \multirow{2}{*}{1.87} &    \multirow{2}{*}{2646$\substack{+345 \\ -309}$}\\
        &  &   & \\  
  \multirow{2}{*}{34.53} &  \multirow{2}{*}{2.17$\substack{+1.72 \\ -1.10}$} & \multirow{2}{*}{2.56} &    \multirow{2}{*}{1209$\substack{+140 \\ -129}$} \\
  &  \\
  & & \multirow{2}{*}{3.63} &    \multirow{2}{*}{552.1$\substack{+47.6 \\ -44.5}$} \\
  &  \\
  & & \multirow{2}{*}{4.96} &    \multirow{2}{*}{238.6$\substack{+31.3 \\ -24.8}$} \\
  &  \\
  & & \multirow{2}{*}{6.06} &    \multirow{2}{*}{155.4$\substack{+18.8 \\ -16.9}$} \\
  &  \\
  & & \multirow{2}{*}{7.85} &    \multirow{2}{*}{54.52$\substack{+21.06 \\ -13.99}$} \\
  &  \\
  & & \multirow{2}{*}{8.93} &    \multirow{2}{*}{24.16$\substack{+6.12 \\ -4.59}$} \\
  &  \\
  & & \multirow{2}{*}{11.14} &    \multirow{2}{*}{12.88$\substack{+7.79 \\ -4.84}$} \\
  &  \\
  & & \multirow{2}{*}{15.52} &    \multirow{2}{*}{5.31$\substack{+3.02 \\ -1.79}$} \\
  &  \\

\hline\hline 
 
\end{tabular}  
\label{table2}
  \end{center}
\end{table}

\begin{table}
\caption{Best-fit Broken Power Laws for the Combined Differential Number Counts at 450 $\mu{\rm m}$ and 850 $\mu{\rm m}$}
\begin{center}
\begin{tabular}{ccccc}
\hline \hline	

 Wavelengths  & $N_0$ & $S_0$ & $\alpha$ & $\beta$   \\
  ($\mu$m)      &   (mJy$^{-1}$~deg$^{-2}$)  &  (mJy) &      &        \\

\hline

\multirow{2}{*}{450}                 & \multirow{2}{*}{33.3$^{+44.0}_{-18.1}$} &  \multirow{2}{*}{20.1$^{+6.8}_{-5.1}$}     &  \multirow{2}{*}{2.34$^{+0.14}_{-0.17}$}   & \multirow{2}{*}{5.06$^{+5.03}_{-1.52}$}  \\ 
 & & & & \\

\multirow{2}{*}{850}                 & \multirow{2}{*}{342$^{+56}_{-80}$}   &  \multirow{2}{*}{4.59$^{+0.26}_{-0.38}$}   &  \multirow{2}{*}{2.12$^{+0.06}_{-0.07}$}   & \multirow{2}{*}{3.73$^{+0.59}_{-0.47}$}   \\  
 & & & & \\
 
\hline\hline 
 
\end{tabular}  
\label{table3}
\end{center}
\end{table}

\subsection{The Effect of Multiplicity}\label{sec:mult}

Semi-analytical simulations have shown that source blending could impact the number counts obtained from single-dish observations \citep{2013MNRAS.434.2572H,Cowley2015}. Some recent studies with ALMA 
observations have also discussed the effect of multiplicity on the number counts \citep{Hodge2013An-ALMA-Survey-,Karim2013An-ALMA-survey-,Simpson2015The-SCUBA-2-Cos}. In \cite{Chen2013Resolving-the-C}, 
we used the SMA detected sample in CDF-N \citep{Barger2014Is-There-a-Maxi} to obtain the multiple fraction as a function of flux at $S_{850\mu \rm{m}} >$ 3.5 mJy, and we computed the multiplicity-corrected 
CDF-N 850 $\mu$m number counts above 3.5 mJy, assuming that all the blends split into two equal components. There is one incorrect coefficient in Equation (4) of \cite{Chen2013Resolving-the-C}, which should instead be written as
\begin{equation}   
\begin{aligned}
 \frac{dN_{{\rm corr}}(S)}{dS} = &~  \frac{dN_{{\rm orig}}(S)}{dS} \times (1-f_{{\rm mul}}(S)) \\
      & + f_{{\rm mul}}(2S) \times 2 \times \frac{dN_{{\rm orig}}(2S)}{dS},
\end{aligned}
\end{equation}   
where $f_{{\rm mul}}$ is the multiple fraction of the SMA detected SCUBA-2 sources as a function of flux, and  ${dN_{{\rm corr}}}/{dS}$ and ${dN_{{\rm orig}}}/{dS}$ are the multiplicity-corrected and the original 
SCUBA-2 counts, respectively\footnote{When the flux ratio of the doublets equals $x /y$, with $x+y=1$, Equation (2) becomes 
$dN_{{\rm corr}}(S)/{dS} = dN_{{\rm orig}}(S)/{dS} \times (1-f_{{\rm mul}}(S)) + f_{{\rm mul}}(S/x) \times  {dN_{{\rm orig}}(S/x)}/{dS} +f_{{\rm mul}}(S/y) \times  {dN_{{\rm orig}}(S/y)}/{dS}$}. 
However, this correction does not significantly change the result of \cite{Chen2013Resolving-the-C}. The systematic changes introduced 
by multiplicity are still smaller than the statistical errors of the counts.

Computing multiplicity corrections is difficult because the multiple fractions at different fluxes are still not well determined. For SCUBA-2 selected sources, \cite{Simpson2015The-SCUBA-2-Cos}
found a multiple fraction of $61^{+19}_{-15}$\% (17 out of 28) at $S_{850\mu \rm{m}} >$ 4 mJy using ALMA, while \cite{Barger2014Is-There-a-Maxi} found a multiple fraction of only $12.5^{+12.1}_{-6.8}$\% (3 out of 24) 
at $S_{850\mu \rm{m}} >$ 3.5 mJy using SMA. The much lower multiple fraction from \cite{Barger2014Is-There-a-Maxi} can be caused by the sensitivities of their SMA maps, which only allow detections of secondary SMGs 
brighter than 3 mJy. The multiple fraction is simply sensitive to the depth of the follow-up interferometric observations. However, \cite{Chen2013Resolving-the-C} showed that most of the sources with a single SMA detection in CDF-N have flux measurements that statistically agree with those made by SCUBA-2. Using a larger SMA detected sample in CDF-N (31 4$\,\sigma$ detected sources; Cowie et al. 2016, in preparation), we again compare the fluxes measured by SCUBA-2 and by SMA for the sources with a single SMA detection. We show the comparison in Figure~4. The two fluxes statistically agree with each other for most of the sources. The median flux ratio of SMA to SCUBA-2 is $0.96 \pm 0.06$. This suggests that most secondary sources that are missed by SMA would be faint and unlikely to affect the bright-end counts. These sources would be unlikely to increase the faint-end counts significantly, either, since they contribute a small fraction of the faint sources based on the broken power law model.

Instead of computing the multiplicity corrections by assuming the multiple fractions, here we use another approach to examine the effect of multiplicity on the number counts. We took the SMA detected sample in CDF-N 
(Cowie et al. 2016, in preparation) and their corresponding SCUBA-2 sources to compute two sets of number counts. We corrected the SCUBA-2 counts for Eddington bias using simulations. For the SMA sources, we 
simply took their fluxes and computed the detectable areas and number counts as if they were detected in our SCUBA-2 map. Note that these sources comprise a incomplete sample at $S_{850\mu \rm{m}} >$ 3 mJy because 
only 29 out of 81 SCUBA-2 sources above this flux (19 out of 29 at $S_{850\mu \rm{m}} >$ 6 mJy) have SMA observations. We did not apply any completeness correction since we are only attempting to see whether there is a
significant difference between these two sets of counts. The comparison is shown in Figure~5. We can see small deviations at both the faint and bright ends, but the two sets of counts are essentially consistent within their uncertainties.

Although \cite{Simpson2015The-SCUBA-2-Cos} found a multiple fraction of 61\% at $S_{850\mu \rm{m}} >$ 4 mJy, their differential counts constructed from ALMA and SCUBA-2 still statistically 
agree with each other at $S_{870\mu \rm{m}}\lesssim$ 15 mJy (see their Figure~6). The effect of multiplicity is more obvious in the cumulative counts at $S_{870\mu \rm{m}} \gtrsim $ 10 mJy. If we plot the cumulative 
EBL, the ALMA and SCUBA-2 results of \cite{Simpson2015The-SCUBA-2-Cos} would deviate at $S_{870\mu \rm{m}} > $ 5 mJy, which is well above the confusion limit of the JCMT. As we will discuss in 
Section~\ref{sec:ebl}, the majority of the EBL comes from sources fainter than our detection limits at both wavelengths, and this conclusion is not affected by multiplicity.

\begin{figure}
\begin{center}    
\includegraphics[width=7.5cm]{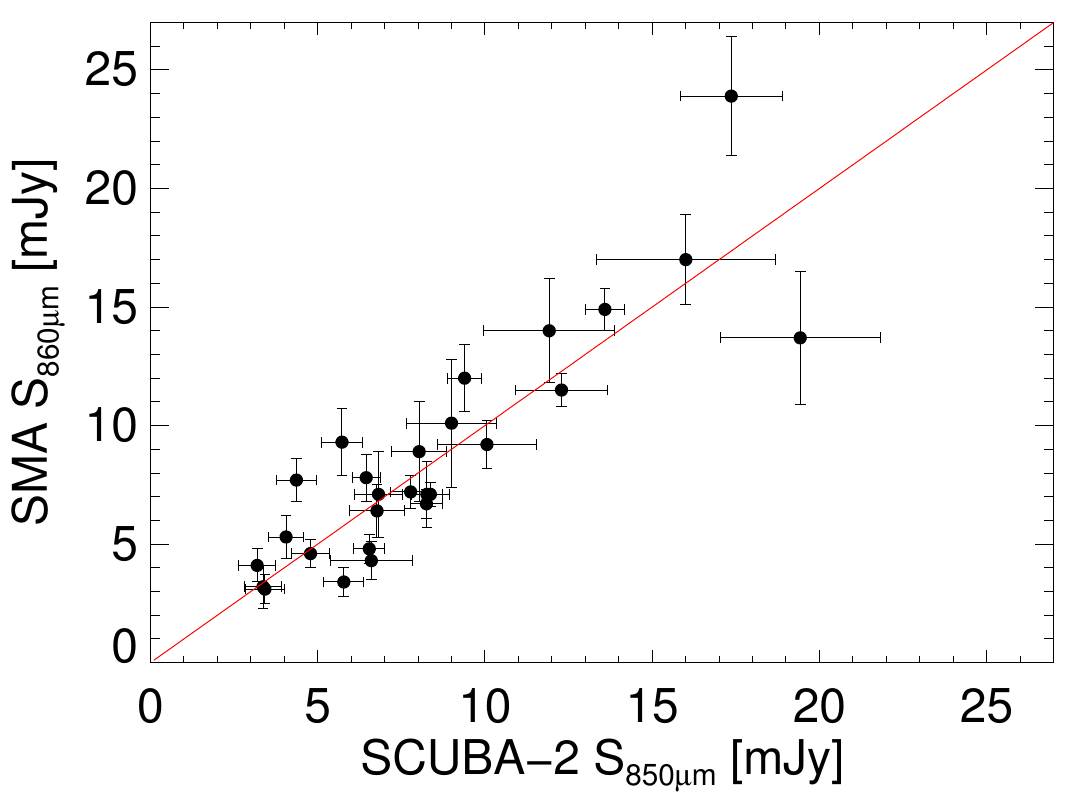}   

\caption{Comparison of SCUBA-2 850 $\mu$m flux and SMA 860 $\mu$m flux for the SCUBA-2 4$\,\sigma$ detected sources in CDF-N with a single SMA detection.}  

\end{center}
\label{figure4}
\end{figure}

\begin{figure}
\begin{center}    
\includegraphics[width=8cm]{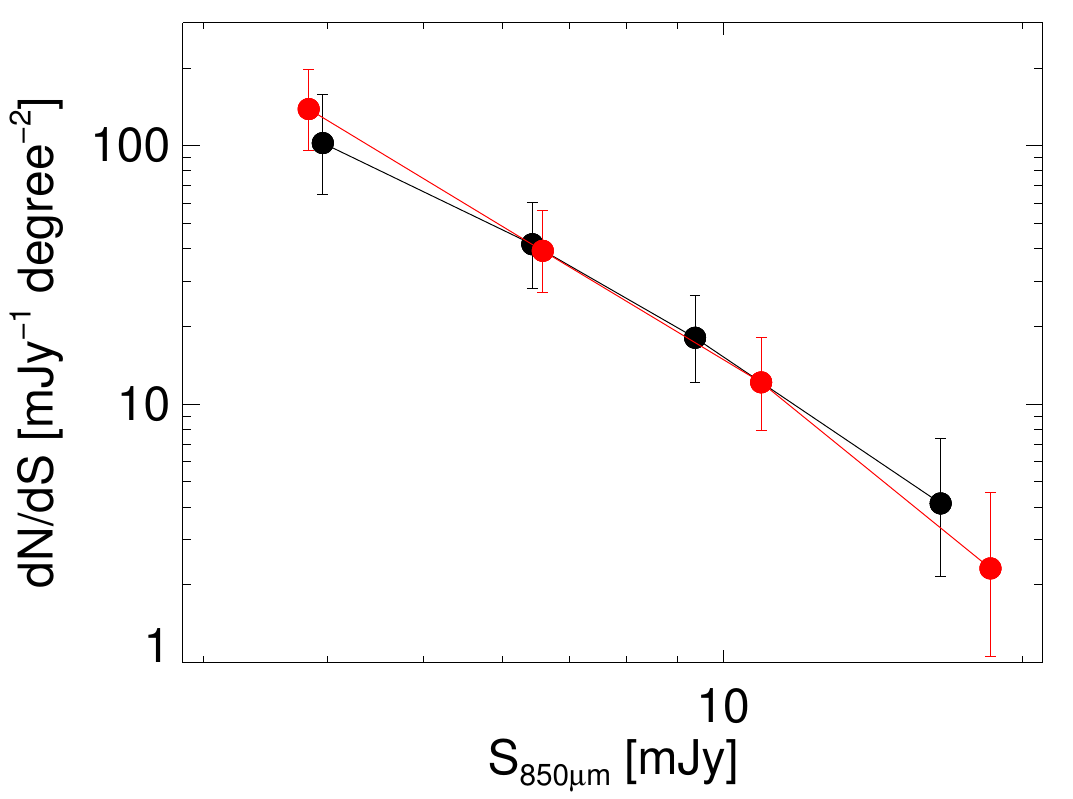}   

\caption{CDF-N 850 $\mu$m differential number counts for the SMA observed SCUBA-2 sources. The black and red circles represent the counts based on SCUBA-2 and SMA fluxes, respectively. These counts are 
lower than the counts for CDF-N shown in Figure~2 because we did not apply any completeness correction. Note that the flux bins for the two sets are different.}  

\end{center}
\label{figure5}
\end{figure}

\section{Discussion}\label{sec:dis}

\begin{figure}
\begin{center}    
\includegraphics[width=8.5cm]{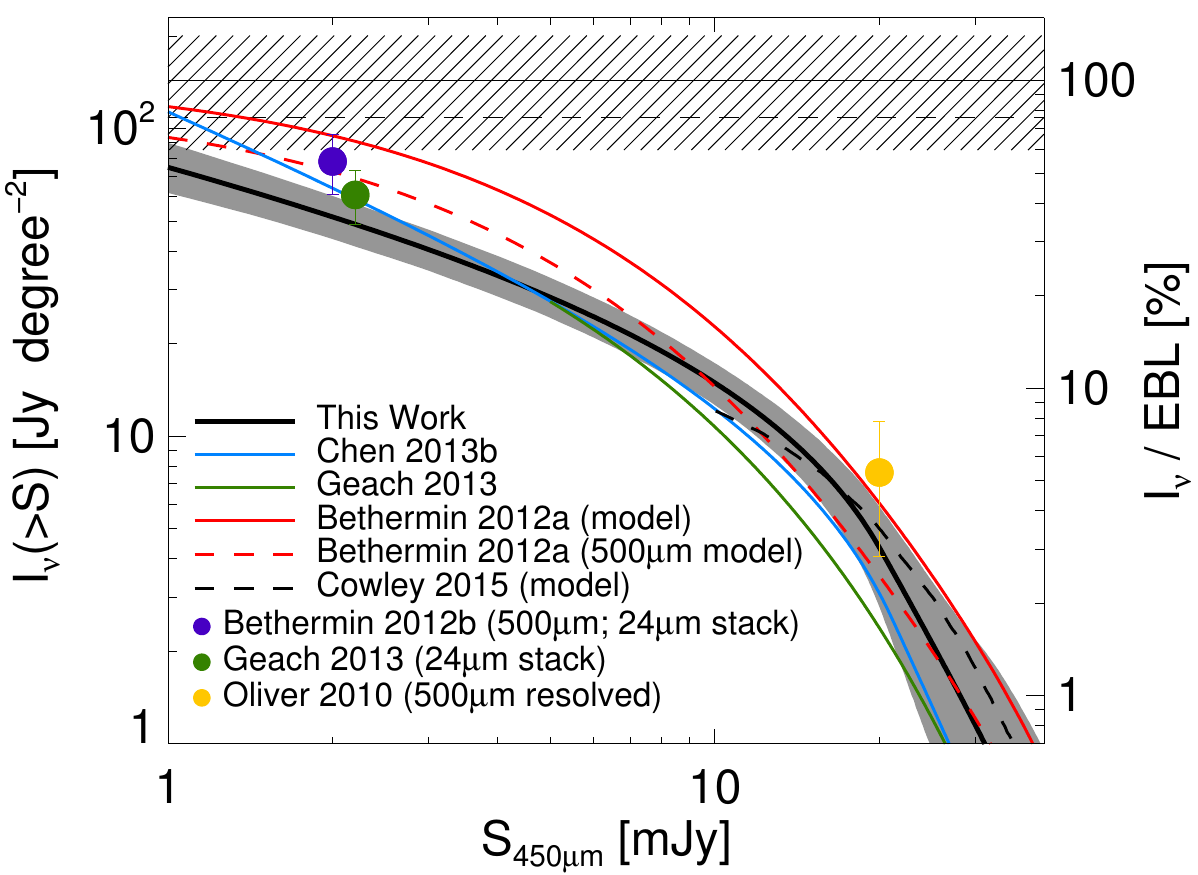}   
\includegraphics[width=8.5cm]{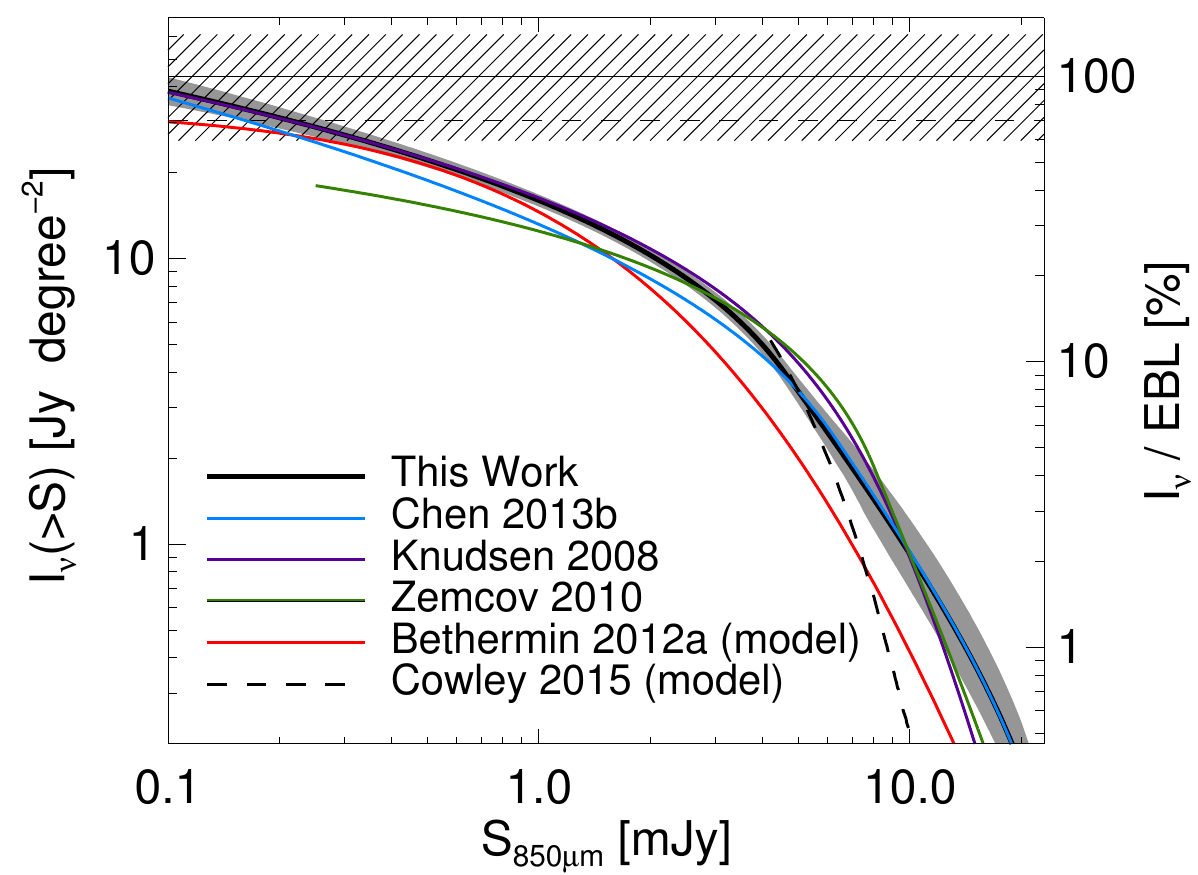}  
\caption{Cumulative EBL as a function of flux at 450 $\mu$m (upper) and 850 $\mu$m (lower). The thick solid black curves are calculated from our best-fit broken power law models (Table~3) with 1\,$\sigma$ error regions in gray shading. The 
horizontal black dashed line \citep{Puget1996Tentative-detec} and the horizontal solid line with the hatched regions \citep{Fixsen1998The-Spectrum-of} are the EBL measured with the {\it COBE} satellite. The blue curves represent our previous results from \cite{Chen2013Resolving-the-C}. The predictions at 450/500 $\mu$m and 850 $\mu$m from \cite{Bethermin2012A-Unified-Empir} are shown as solid and dashed red curves. The black dashed curves are the predictions from \cite{Cowley2015}. The green curve in the upper panel represents the result of the SCUBA-2 450 $\mu$m map of the COSMOS field from \cite{2013MNRAS.432...53G}. The purple and green curves in the lower panel are the results based on SCUBA cluster surveys from \cite{Knudsen2008Probing-the-sub} and \cite{Zemcov2010Contribution-of}, respectively. In the upper panel, some results of the 24 $\mu$m stacking on the 450 $\mu$m \citep{2013MNRAS.432...53G} and 500 $\mu$m \citep{Bethermin2012HerMES:-deep-nu} maps and the directly resolved 500 $\mu$m background light \citep{Oliver2010HerMES:-SPIRE-g} are shown as colored circles. }  

\end{center}
\label{figure6}
\end{figure}

\begin{figure}
\begin{center}    
\includegraphics[width=7.5cm]{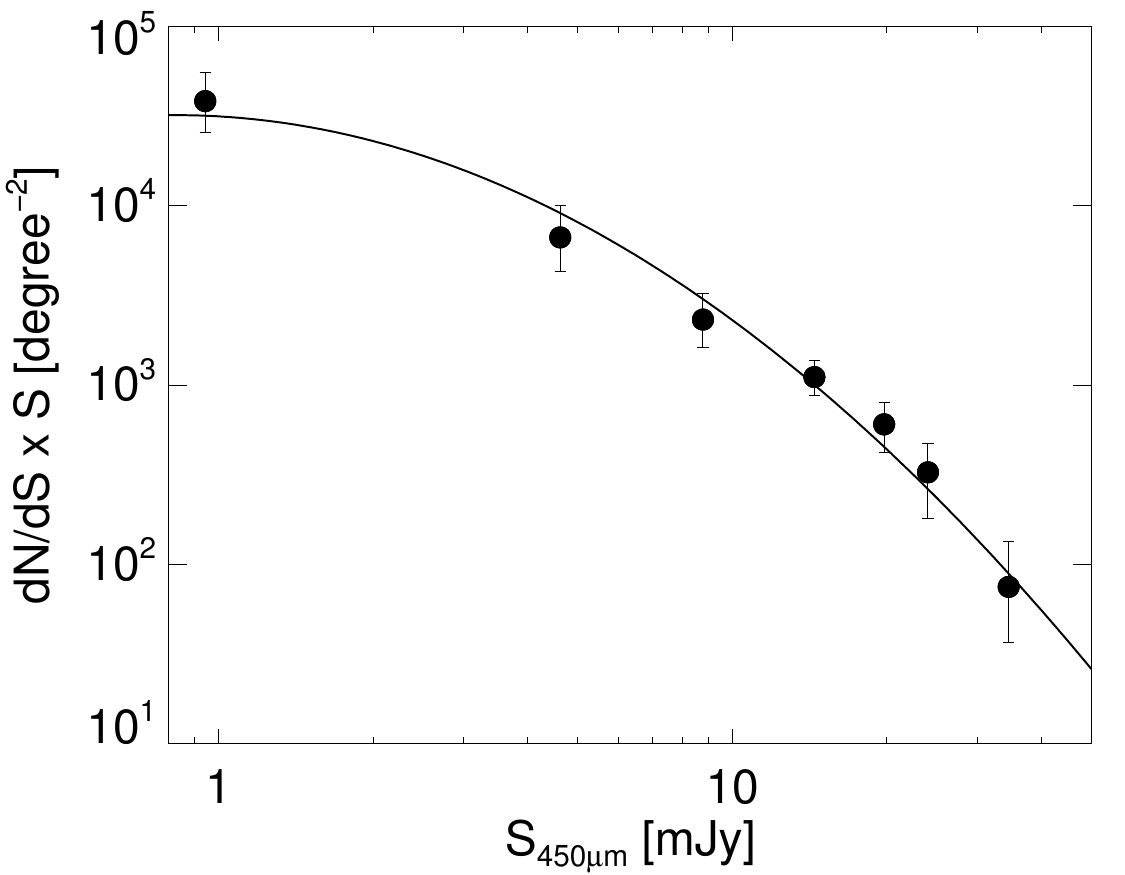} 
\includegraphics[width=7.5cm]{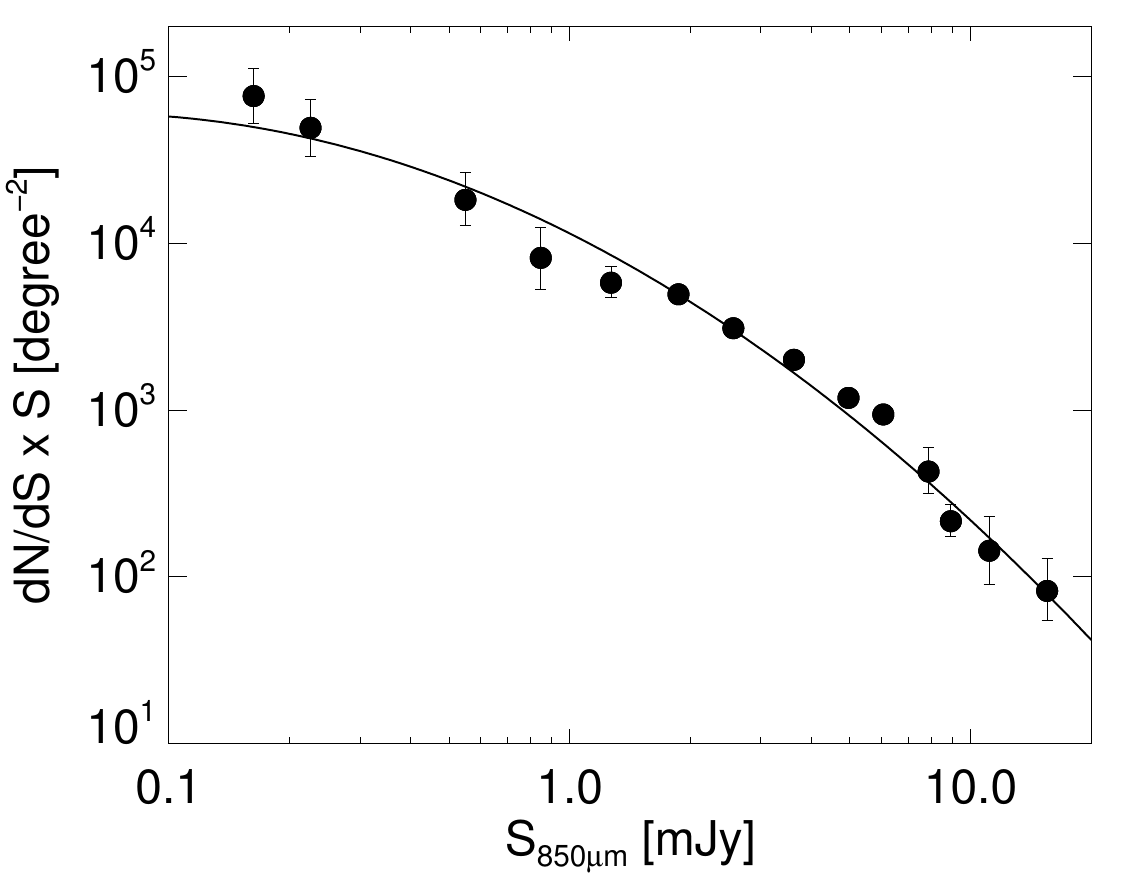}  
\caption{Combined differential number counts (from Figure~3) multiplied by the flux. Because the total EBL is essentially the integral of such curves, the curve for each wavelength must turn over at some point such that the derived cumulative 
EBL does not significantly exceed the total EBL measured by the {\it COBE} satellite. Solid black curves are second order polynomial fits to the log-log plots. The estimated turnovers are at $\sim$ 0.8 mJy at 450 $\mu$m and $\sim$ 0.06 mJy 
at 850 $\mu$m based on the polynomial fits.}   
\end{center}
\label{figure7}
\end{figure}

\begin{figure*}
\begin{center}    
\includegraphics[width=6.5cm]{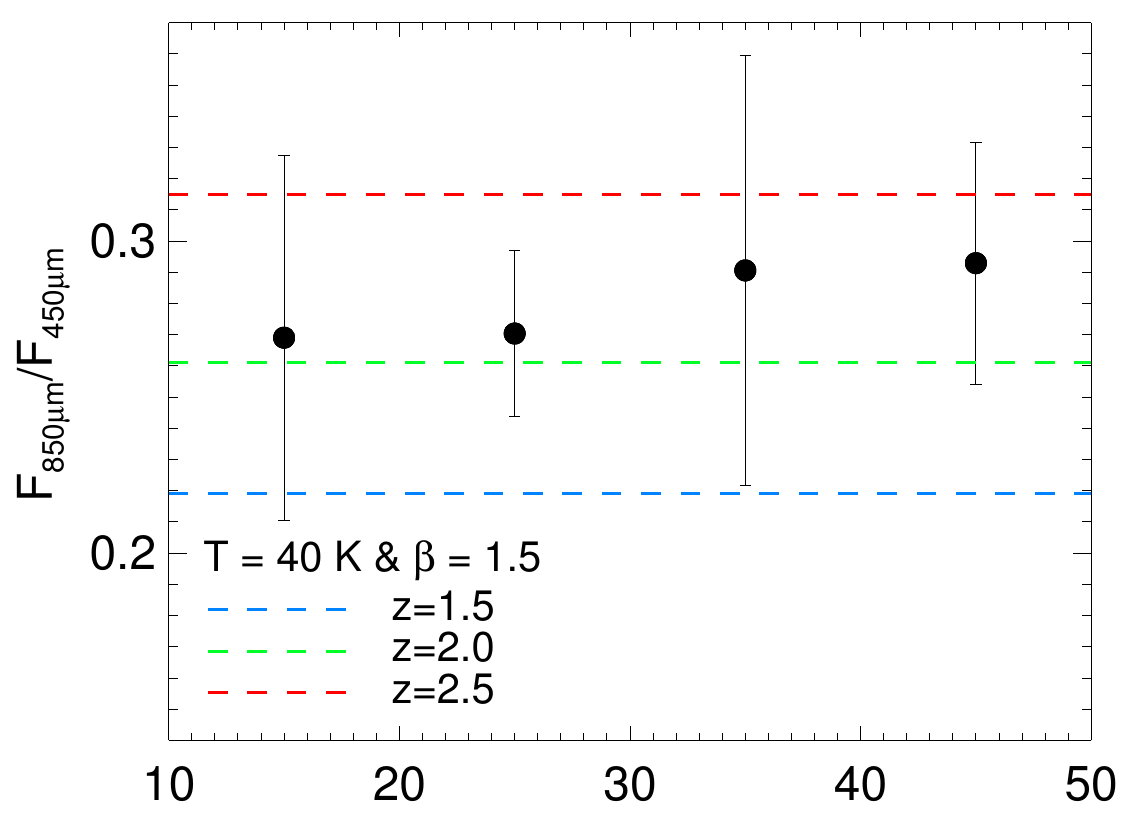}   
\includegraphics[width=6.5cm]{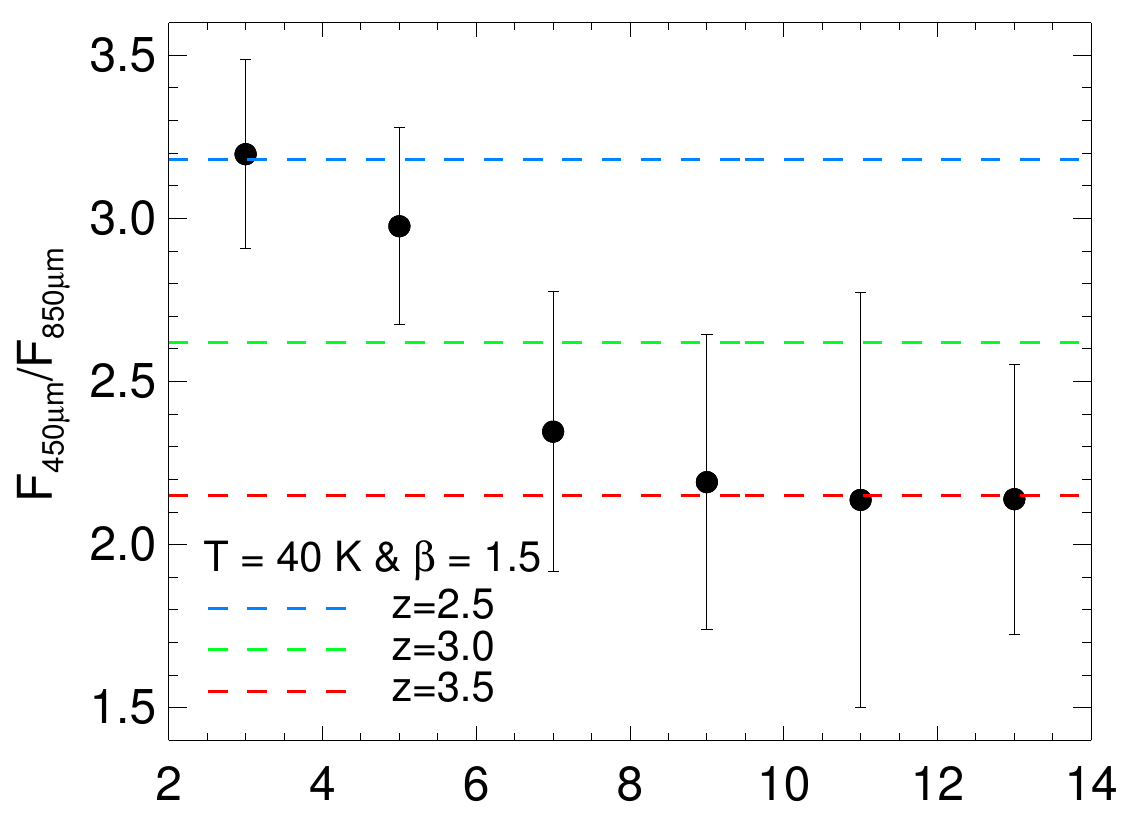}\\   
\includegraphics[width=6.5cm]{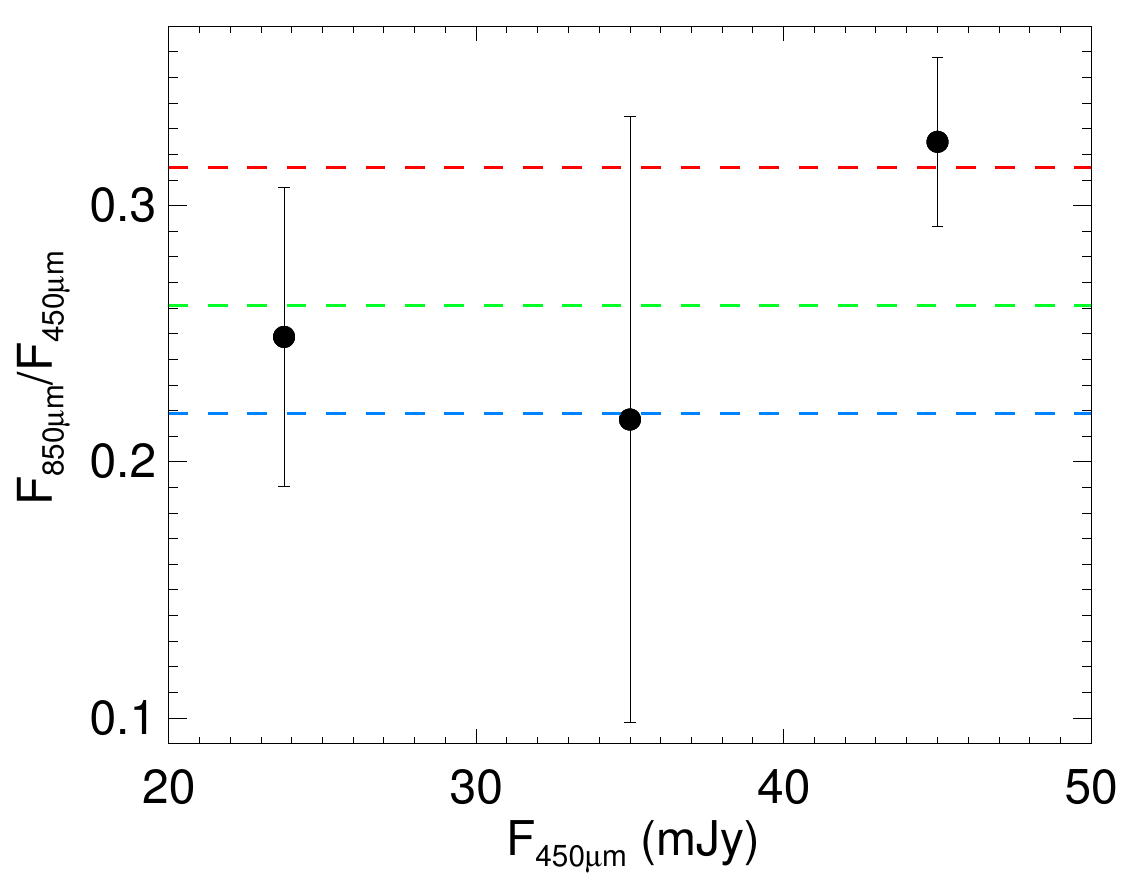}   
\includegraphics[width=6.5cm]{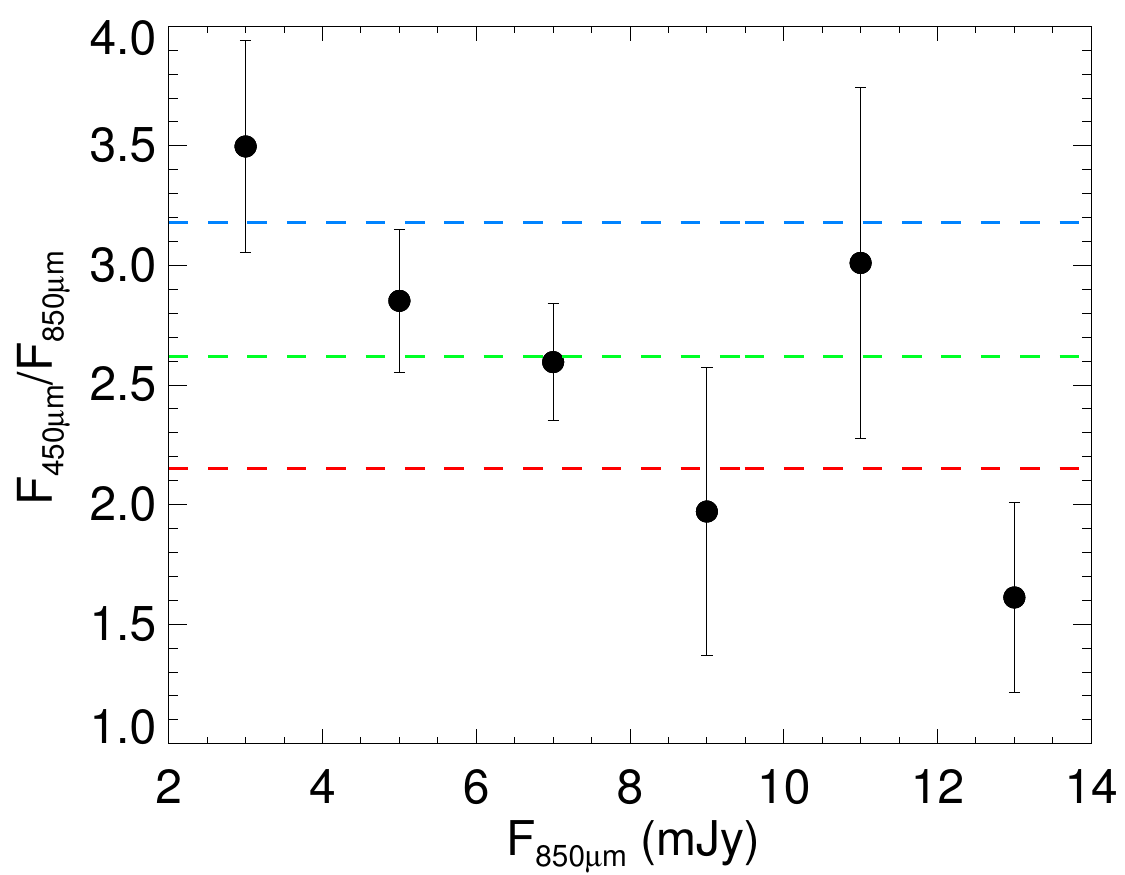}   
\caption{850$\mu$m-to-450$\mu$m flux ratio as a function of 450 $\mu$m flux (left) and 450$\mu$m-to-850$\mu$m flux ratio as a function of 850 $\mu$m flux (right) for 4$\,\sigma$ detected sources from 
the 450 $\mu$m and 850 $\mu$m maps, respectively, for our cluster fields (upper) and blank fields (lower). In each flux bin, we took the median of the flux ratios and calculated the error using bootstrapping. Unlike Figure~1, the 
data points are corrected for the effect of image noises on the flux ratio measurements, which is done by running simulations in which we populated sources with constant flux ratios. Some predicted flux ratios of a modified blackbody 
SED with $\beta = 1.5$ and a dust temperature of 40 K at several redshifts are shown as colored dashed lines for comparison.}  
\end{center}
\label{figure8}
\end{figure*}

\subsection{Extragalactic Background Light}\label{sec:ebl}

We plot the cumulative EBL as a function of flux at both wavelengths in Figure~6. Without gravitational lensing, our surveys would be limited to sources with $S_{450\mu \rm{m}} \gtrsim $ 10 mJy or $S_{850\mu \rm{m}} > $ 2 mJy. However, sources fainter than these limits contribute the majority of the EBL at both wavelengths. If we use the measurement by \cite{Fixsen1998The-Spectrum-of} 
as the total EBL, $\sim$ 90\% of the 450 $\mu{\rm m}$ background comes from sources fainter than 10 mJy and $\sim$ 80\% of the 850 $\mu{\rm m}$ background comes from sources fainter than 2 mJy. These numbers would not 
change significantly even if we consider the effect of multiplicity. Our result also suggests there is at least $\sim$ 50\% of the EBL with $S_{450\mu \rm{m}} <$ 1.0 mJy. If we integrate the 450 $\mu{\rm m}$ differential count down to the lower flux limit in the upper panel of Figure~2, there is still at least $\sim$ 40\% of the EBL with $S_{450\mu \rm{m}} <$ 0.7 mJy that is not resolved by our SCUBA-2 maps. Most of these faint SMGs would have $L_{\rm IR} < 10^{12} L_{\odot }$, corresponding to LIRGs or normal galaxies. However, note that all these fractions of the EBL we calculate here depend on which measurement of the total EBL we assume, as well as its uncertainty.

We note that the faint-end slopes ($\alpha$) of the number counts should become less than one at fluxes fainter than 1 mJy at 450 $\mu$m and 0.1 mJy at 850 $\mu$m. In Figure~7, we show the combined differential 
number counts (from Figure~3) multiplied by the flux. Because the total EBL equals the integral of $dN/dS \times S$ over $S$, $dN/dS \times S$ must turn over at some point such that the derived cumulative 
EBL does not significantly exceed the total EBL measured by the {\it COBE} satellite. Using second order polynomial fits to the log-log plots in Figure~7, the estimated turnovers (where $\alpha$ becomes one) are at 
$\sim$ 0.8 mJy at 450 $\mu$m and $\sim$ 0.06 mJy at 850 $\mu$m. Sources with these fluxes would contribute the most to the EBL. If we again assume a modified blackbody SED with $\beta = 1.5$ and a 
dust temperature of 40 K, $S_{450\mu \rm{m}} \sim$ 0.8 mJy and $S_{850\mu \rm{m}} \sim$ 0.06 mJy correspond to $L_{\rm IR} \sim 1.3 \times 10^{11} L_{\odot}$ at $z = 2.2$ and $\sim 3.4 \times 10^{10} L_{\odot}$ at $z = 2.8$, respectively.

We also show other measurements and model predictions of the EBL from the literature in Figure~6. All of the observational results are consistent with ours within 1\,$\sigma$ (note that the 1\,$\sigma$ spread of all the 
other cumulative EBL curves are not shown in Figure~6). There is, however, significant disagreement between the 450 $\mu$m model from \cite{Bethermin2012A-Unified-Empir} and our result. This difference is mainly 
caused by the discrepancy in the differential number counts at $S_{450\mu \rm{m}} \sim$ 2--15 mJy (see Figure~3), where the model slightly overproduces sources.

\cite{Viero2013HerMES:-The-Con} quantified the fraction of the EBL that originates from galaxies identified in the UV/optical/near-infrared by stacking $K$-selected sources on various {\it Spitzer} and {\it Herschel} maps at different 
wavelengths. They were able to resolve 65\%$\pm$12\% of the EBL at 500 $\mu$m (2.60 nW m$^{-2}$ sr$^{-1}$ or 0.434 MJy sr$^{-1}$) based on the measurement by \cite{Lagache2000Evidence-for-du}. If we correct their result using the EBL measured by \cite{Fixsen1998The-Spectrum-of}, their sample contributes $\sim$~70\% of the EBL at 500 $\mu$m (2.37 nW m$^{-2}$ sr$^{-1}$ or 0.395 MJy sr$^{-1}$), which includes $\sim$~10\%, 40\%, and 20\% coming from normal galaxies, LIRGs, and ULIRGs, respectively. For comparison, we can assume an extreme case, where all of the 450 $\mu$m sources lie at $z = 2.8$ and have a modified blackbody SED with $\beta = 1.5$ and $T = $ 50 K (see Section~\ref{sec:color}). In such a case, galaxies with $L_{\rm IR} < 10^{12} L_{\odot }$ would have $S_{450\mu \rm{m}} \lesssim $ 3.4 mJy, which still contribute $\sim$ 75 \% of the EBL and cannot be fully accounted for by the normal galaxies and LIRGs in \cite{Viero2013HerMES:-The-Con}. If we assume a lower dust temperature or a lower redshift, galaxies with $L_{\rm IR} < 10^{12} L_{\odot }$ would contribute even more to the EBL. This is consistent with recent SMA \citep{Chen2014SMA-Observation} and ALMA \citep{Kohno2016SXDF-UDS-CANDEL,Fujimoto2016ALMA-Census-of-} observations, which suggest that many faint SMGs may not be included in the UV star formation history.

Because the majority of sources that contribute the submillimeter EBL have $L_{\rm IR} < 10^{12} L_{\odot }$, a full accounting of the cosmic star formation history requires a thorough understanding of the galaxies with FIR luminosities corresponding to LIRGs and normal star-forming galaxies. It is therefore critical to determine how much the submillimeter- and UV-selected samples overlap at this luminosity range. Future work using interferometry is needed to determine the fraction of faint SMGs that is already included in the UV population as a function of submillimeter flux.

\subsection{Redshift Distributions}\label{sec:redshift}

As described in Section~\ref{sec:color}, we used a statistical approach to explore the submillimeter flux ratios for both 450 $\mu$m and 850 $\mu$m selected samples from the cluster fields. If we do the same 
exercise on the blank-field data and again use a modified blackbody SED with $\beta = 1.5$, $T=$ 40 K, the median redshifts of the 450 $\mu$m and 850 $\mu$m populations would be 2.0 and 
2.6, respectively. These are in rough agreement with the median redshifts (2.06 $\pm$ 0.10 and 2.43 $\pm$ 0.12) from the simulations of \cite{2014MNRAS.443.2384Z}. \cite{Bethermin2015The-influence-o} 
also presented the median redshift of dusty galaxies as a function of wavelength, flux limit, and lensing selection bias based on their empirical model \citep{Bethermin2012A-Unified-Empir}. The 4\,$\sigma$ detection 
limits for our blank-field 450 $\mu$m, cluster 450 $\mu$m, blank-field 850 $\mu$m, and cluster 850 $\mu$m images are $\sim$ 18, 10, 2, and 2 mJy, respectively. According to Figure~3 of \cite{Bethermin2015The-influence-o}, 
these flux cuts correspond to median redshifts of $z \sim$ 1.9, 1.8 $\lesssim z \lesssim$ 2.4, $z \sim$ 2.4 and 2.4 $\lesssim z \lesssim$ 2.8, respectively. These also roughly agree with our estimated median redshifts. Note 
that for our cluster fields, the corresponding redshifts are shown as intervals, because the relations in Figure~3 of \cite{Bethermin2015The-influence-o} are for all galaxies and ``strongly lensed'' galaxies, while our SCUBA-2 
maps extend out to radii of $\sim$ 6$'$ and therefore detect both strongly and weakly lensed sources.

In Figure~8, we show the 850$\mu$m-to-450$\mu$m and 450$\mu$m-to-850$\mu$m flux ratios versus the observed 450 $\mu$m and 850 $\mu$m fluxes, respectively, for both our cluster and blank-field data. The 
main difference between Figures~1 and 8 is that we correct the data points in Figure~8 for the effect of image noise on the flux ratio measurements. This is again done by running simulations in which we generated sources with constant flux ratios, and a detailed description is left to the Appendix. In Figure~8, we also show some predicted flux ratios of a modified blackbody SED with $\beta = 1.5$ and a dust temperature of 40 K at several redshifts. At 850 $\mu$m, a clear relation between the flux ratio and the observed flux can be seen in both the cluster and blank fields. This relation can be explained by a redshift evolution if the SEDs of these galaxies do not change significantly with redshift. Since the observed 850 $\mu$m flux of an SMG remains almost invariant over $z=$ 1--8 due to the strong negative {\it K}-correction, the variation of the flux ratio we see here might be a result of ``cosmic downsizing'' \citep{Cowie1996New-Insight-on-}, where SMGs at higher redshifts have higher gas fractions and therefore higher luminosities and star formation rates (e.g., \citealt{Heavens2004The-star-format,Bundy2006The-Mass-Assemb,Franceschini2006Cosmic-evolutio,Dye2008The-SCUBA-HAlf-,Mobasher2009Relation-Betwee,Magliocchetti2011The-PEP-survey:}). Note that although the variation can be explained by an evolution in dust temperature, it would be interpreted as brighter sources having lower temperatures, which conflicts with many studies of dusty star-forming galaxies (e.g., \citealt{Casey2012A-Population-of,U2012Spectral-Energy,Lee2013Multi-wavelengt,Symeonidis2013}).  

Another possible factor that can cause the redshift variation we see here would be lensing bias, in which brighter sources contain a higher fraction of high-redshift, lensed galaxies. For the sources in our cluster fields, although their 
redshifts are needed to compute the precise lensing magnifications, the changes in their magnifications are much more sensitive to the source positions than to the redshifts. As a consequence, we can use the magnification maps for $z=2.2$ and $z=2.8$ that we generated using {\sc LENSTOOL} (see Section~\ref{sec:delensing}) to quantify how strong the lensing effect is for each source. We do not see any correlation between the observed flux and the magnification for 
our lensed sources. This suggests that these brighter sources in the cluster fields are generally not more strongly lensed and are simply brighter intrinsically.

We also cannot rule out the possibility of galaxy--galaxy strong lensing events (which are not included in the cluster lens models) that cause lensing bias in both the blank and cluster fields. Theoretical predictions (e.g., \citealt{Blain1996Galaxy-galaxy-g,Perrotta2002Gravitational-l,Perrotta2003Predictions-for,Negrello2007Astrophysical-a,Paciga2009Strong-lensing-,Bethermin2012A-Unified-Empir,Wardlow2013HerMES:-Candida}) showed that the fraction of strongly lensed sources increases with the observed submillimeter flux. Wide-area, flux-density limited surveys with {\it Herschel} have successfully discovered many bright, strongly lensed SMGs (e.g., \citealt{Negrello2010The-Detection-o,Conley2011Discovery-of-a-,Wardlow2013HerMES:-Candida}). However, at the flux range of our SCUBA-2 sources, these galaxy--galaxy strong lensing events should be rare and should have little effect on the observed redshift distribution. If we take the count models from \cite{Bethermin2012A-Unified-Empir}, 850 $\mu$m sources with fluxes of 3, 5, 7, 9,11, and 13 mJy (corresponding to the flux bins in Figure~8) have lensing fractions of $\sim$ 1\%, 2\%, 3\%, 5\%, 7\%, and 10\%, respectively. Although a fraction of 10\% might cause a significant effect, the lensing fractions in the four faintest 850 $\mu$m flux bins in Figure~8 are too small to produce the variation of redshift we see in both cluster and blank fields. Therefore, we conclude that lensing bias only has a minor effect on the observed redshift distribution and a downsizing scenario is the most likely cause. 

On the other hand, we do not see a clear relation between the submillimeter flux ratio and the 450 $\mu$m flux, mainly because of the large uncertainties due to small number statistics. Deeper 450 $\mu$m maps obtained in the future 
should improve our measurements. However, a nearly flat distribution for the lensed sources shown in Figure~8 is in agreement with \cite{2013MNRAS.436..430R}. This result might still be consistent with a downsizing scenario, given that the observed 450 $\mu$m flux of an SMG does decrease with the redshift. The observed distribution of flux ratios might be flattened due to a mixture of high-redshift bright and low-redshift faint objects in the same flux bin. A similar trend is also seen in Figure~3 of \cite{Bethermin2015The-influence-o}, where the median redshift of 450 $\mu$m sources between flux-density cuts of 10 and 50 mJy (the flux range shown in Figure~8) changes less than that of 850 $\mu$m sources between flux-density cuts of 2 and 14 mJy (the flux range shown in Figure~8) in both full samples and strongly lensed samples. Again, if we take the count models from \cite{Bethermin2012A-Unified-Empir}, 450 $\mu$m sources with fluxes of 15, 25, 35, and 45 mJy (corresponding to the flux bins in Figure~8) have galaxy--galaxy lensing fractions of $\sim$ 1\%, 2\%, 4\%, and 7\%, respectively, which have little effect on the observed redshift distribution. 

\section{Summary}\label{sec:sum}

Using the SCUBA-2 camera mounted on the JCMT, we present deep number counts at 450 and 850 $\mu$m. We combine data of six lensing cluster fields (A1689, A2390, A370, MACS\,J0717.5+3745, MACS\,J1149.5+2223, and MACS\,J1423.8+2404) and three blank fields (CDF-N, CDF-S, and COSMOS) to measure the counts over a wide flux range at each wavelength. Thanks to gravitational lensing, we are able to detect counts at fluxes fainter than 1 mJy and 0.2 mJy in several fields at 450 $\mu$m and 850 $\mu$m, respectively. With the large number of cluster fields, our combined data highly constrain the faint end of the number counts. By integrating the number counts we measure, we found that the majority of EBL at each wavelength is contributed by sources that are fainter than the detection limit of our blank-field images. Most of these faint sourcess would have $L_{\rm IR} < 10^{12} L_{\odot }$, corresponding to LIRGs or normal galaxies. By comparing our result 
with the 500 $\mu$m stacking of $K$-selected sources from \cite{Viero2013HerMES:-The-Con}, we conclude that the $K$-selected LIRGs and normal galaxies still cannot fully account for the EBL that originates from sources with $L_{\rm IR} < 10^{12} L_{\odot }$. This is consistent with recent SMA \citep{Chen2014SMA-Observation} and ALMA \citep{Kohno2016SXDF-UDS-CANDEL} observations, which suggest that many faint SMGs may not be included in the UV star formation history. We also explore the submillimeter flux ratio between the two wavelengths for our 450 $\mu$m and 850 $\mu$m selected sources. At 850 $\mu$m, we find a clear relation between the flux ratio with the observed flux. This relation can be explained by a redshift evolution if the SEDs of these SMGs do not change significantly with redshift, where galaxies at higher redshifts have higher luminosities and star formation rates. On the other hand, we do not see a clear relation between the flux ratio and 
450 $\mu$m flux.

\acknowledgments

We thank the anonymous referee for a careful and thoughtful reading of the original version of this paper and for offering suggestions to improve both its substance and presentation. We gratefully acknowledge 
support from NSF grants AST-1313309 (L.-Y.H., L.L.C.) and AST-1313150 (A.J.B.), the ERC Advanced Investigator program DUSTYGAL 321334 (C.-C.C.), the University of Wisconsin Research Committee with funds 
granted by the Wisconsin Alumni Research Foundation (A.J.B.), and the William F. Vilas Trust Estate (A.J.B.). We also thank JCMT support astronomers Iain Coulson and Jan Wouterloot, and JCMT telescope operators Jim Hoge, Callie McNew and William Montgomerie. The James Clerk Maxwell Telescope is operated by the East Asian Observatory on behalf of The National Astronomical Observatory of Japan, Academia Sinica Institute of Astronomy and Astrophysics, the Korea Astronomy and Space Science Institute, the National Astronomical Observatories of China and the Chinese Academy of Sciences (Grant No. XDB09000000), with additional funding support from the Science and Technology Facilities Council of the United Kingdom and participating universities in the United Kingdom and Canada. The James Clerk Maxwell Telescope has historically been operated by the Joint Astronomy Centre on behalf of the Science and Technology Facilities Council of the United Kingdom, the National Research Council of Canada and the Netherlands Organisation for Scientific Research. Additional funds for the construction of SCUBA-2 were provided by the Canada Foundation for Innovation.

\appendix

As we described in Section~\ref{sec:color}, we compared the measured submillimeter flux ratios with what we measured from the simulated maps for the cluster fields. In these simulations, we populated the pure noise maps with sources with constant flux ratios. In order to simulate maps for a cluster field, we need the underlying count model (de-lensed and Eddington-bias-corrected) to place sources on the source plane and then project them onto the image plane using {\sc LENSTOOL}. However, what we tried to find out is the redshift of the source plane, which is also needed for measuring the de-lensed, corrected count model. Therefore, we measured the number counts and ran simulations in the following iterative way. We first located source planes at $z = 1.4$ at 450 $\mu$m \citep{2013MNRAS.436..430R} and $z = 3.0$ at 850 $\mu$m \citep{Barger2012Precise-Identif,Barger2014Is-There-a-Maxi,2013MNRAS.428.2529H,Vieira2013Dusty-starburst} to measure the number counts of all the cluster fields using the procedure we describe in Sections~\ref{sec:delensing} and \ref{sec:simulation} (which also involved simulations). We then used the final count models we obtained and the assumed source plane redshifts to run simulations on the pure noise maps. To simulate the flux ratio measurement of 450 (850) $\mu$m selected sources, we generated sources on a source plane at $z=1.4$ ($z=3.0$) using the count model we measured, projected them onto the 450 (850) $\mu$m pure noise map at the image plane using {\sc LENSTOOL}, and computed the flux of their 850 (450) $\mu$m counterparts based on the source plane redshift and a modified blackbody SED with $\beta = 1.5$, $T=$ 40 K. We then populated the 850 (450) $\mu$m pure noise map with these counterparts (which again includes lensing projection), and finally, we added additional sources onto the map until it matched the 850 (450) $\mu$m count model.

 With all the simulated maps, we could measure the 850$\mu$m-to-450$\mu$m and 450$\mu$m-to-850$\mu$m flux ratios in the way we describe in Section~\ref{sec:color}. By doing the exercise we show in Figure~1, we found that our simulated 450 $\mu$m and 850 $\mu$m sources need to be redder and slightly bluer, respectively, to match what we measured from the real science maps. We therefore did a second iteration of the whole procedures above, with new input flux ratios and their corresponding source plane redshifts. It took us only three iterations to converge our source plane redshifts to $z = 2.2$ at 450 $\mu$m and $z = 2.8$ at 850 $\mu$m. We also used a dust temperature of 30 K or 50 K to run all
 the procedure above, which yielded to different source plane redshifts but still the same input flux ratios. With $T=$ 30 K, the redshifts are $z \sim$ 1.5 (450 $\mu$m) and $z \sim$ 2.0 (850 $\mu$m); with $T=$ 50 K, the redshifts 
 are $z \sim$ 2.8 (450 $\mu$m) and $z \sim$ 3.5 (850 $\mu$m). In our simulations, although the spatial distribution of the lensed sources and the de-lensed, corrected count models are determined by the source plane redshift we 
 used, it does not influence the flux ratio we measured. In other words, only the input constant flux ratio influence the result, even though we convert it to a redshift using a modified black body SED. In summary, Figures~1 and 2 were made after iterations of all the procedures we describe in Sections~\ref{sec:color}, \ref{sec:delensing} and \ref{sec:simulation}.
 
In order to see how the image noise affect the flux ratio measurements in Figure~8, we again generated sources with constant flux ratios in our simulations. Figure~A1 shows some examples for our cluster fields. We note that 
the measured flux ratios from the simulated maps are not constant as a function of the observed flux. We also found that the ratio between the input, constant value and the recovered value at a certain flux is rather independent 
of the input value. Therefore, we ran multiple simulations with different input constant values and calculated the averaged ratio between the input and recovered values as a function of the flux. We then used the ratio to correct our measurements from the real science maps. The final corrected plots are what we show in Figure~8.

\renewcommand{\thefigure}{A1}

\begin{figure}
\begin{center}    
\includegraphics[width=6cm]{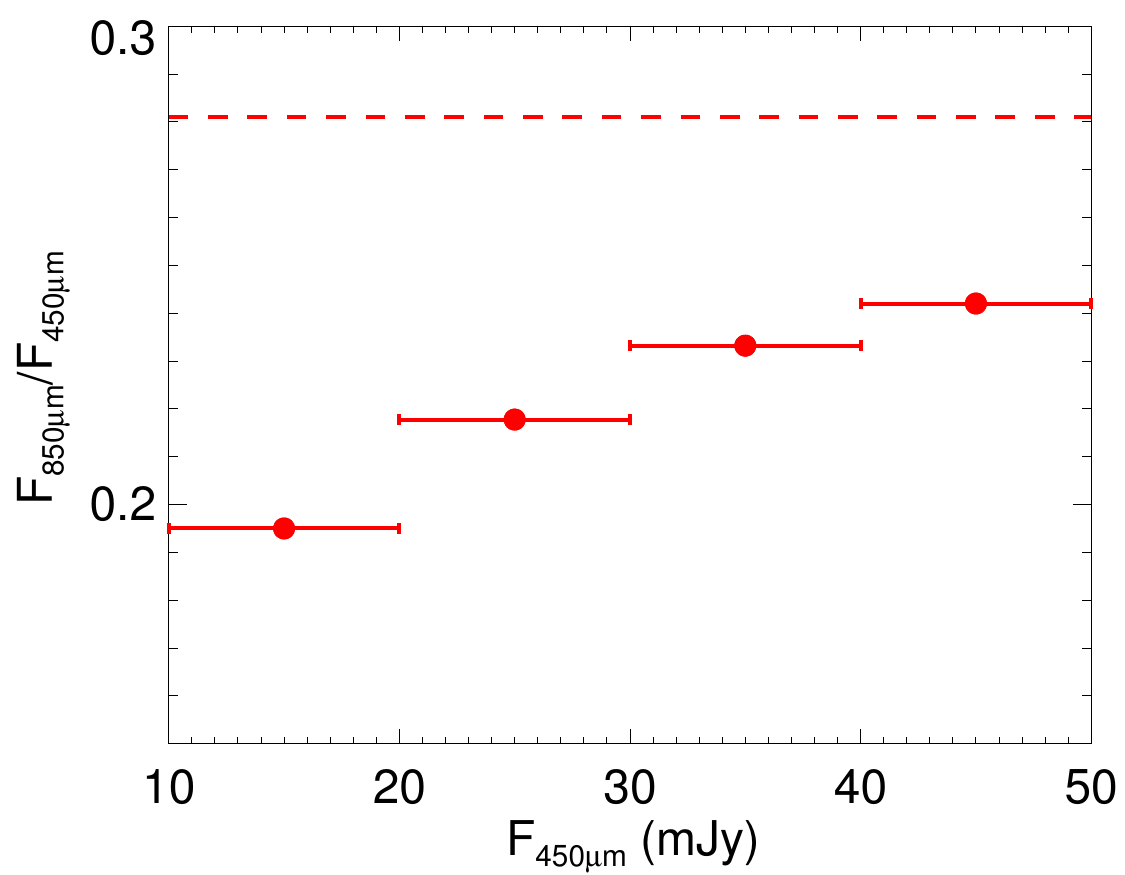}   
\includegraphics[width=6cm]{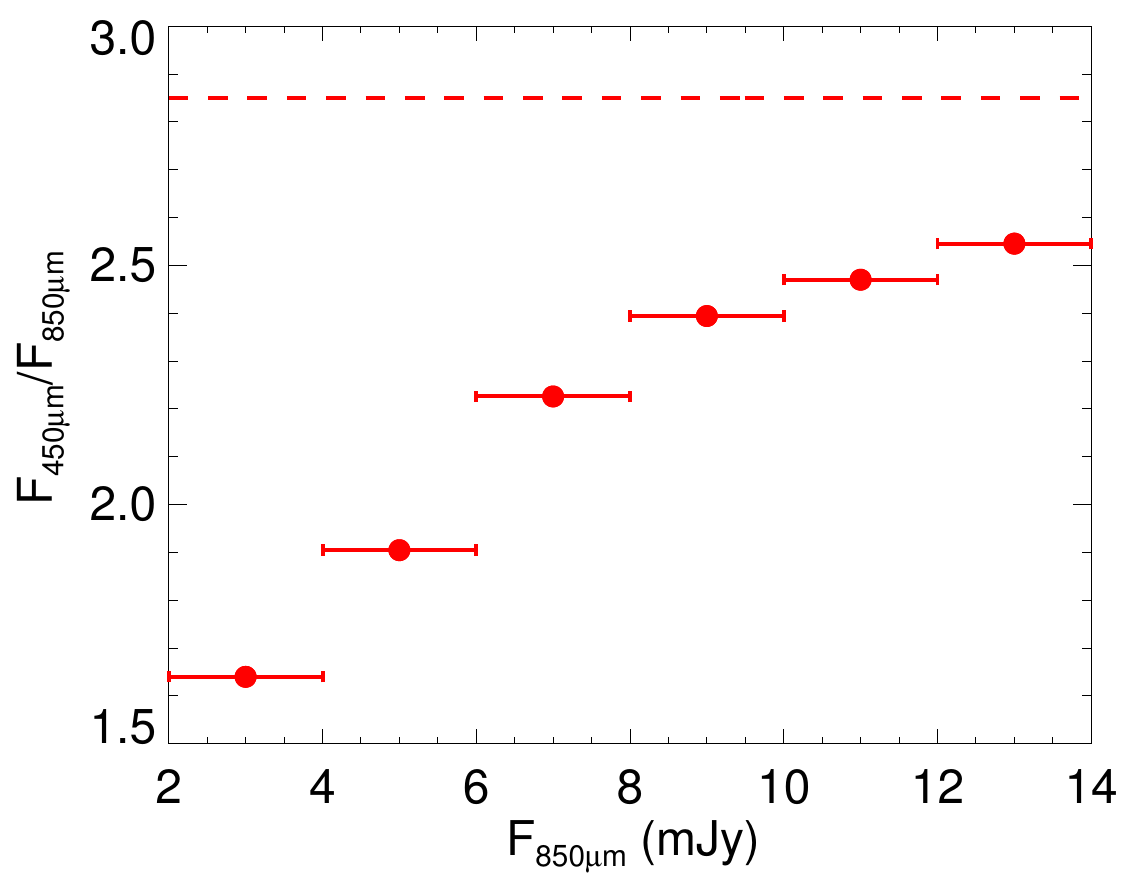}   
\caption{Input (dashed lines) and recovered median (filled circles) submillimeter flux ratios against the observed flux in our simulations for the 450 $\mu$m (left) and 850 $\mu$m selected sources (right) in our cluster fields. Here the input 
values are 0.281 and 2.85, which correspond to $z=2.2$ and $z=2.8$ for the 450 $\mu$m and 850 $\mu$m populations, respectively, based on a modified blackbody SED with $\beta = 1.5$, $T=$ 40 K. The recovered flux ratio is 
measured using the method we describe in Section~\ref{sec:color}.}
\end{center}
\label{figureA}
\end{figure}

% =================== B I B L I O G R A P H Y ========================= %
\bibliographystyle{apj}
\bibliography{ms_revised}

\end{document}